\documentclass[11pt,a4paper]{article}

\usepackage{arxiv} 

\usepackage[T1]{fontenc}
\usepackage[utf8]{inputenc}
\usepackage{authblk}

\author[1]{Alexander Partin, \thanks{Correspondence: apartin@anl.gov}}
\author[1]{Thomas S. Brettin}
\author[1]{Yitan Zhu}
\author[1]{Oleksandr Narykov}
\author[1,2]{Austin Clyde}
\author[1]{Jamie Overbeek}
\author[1,2]{Rick L. Stevens}

\affil[1]{Division of Data Science and Learning, Argonne National Laboratory, Lemont, IL, USA}
\affil[2]{Department of Computer Science, The University of Chicago, Chicago, IL, USA}

\title{Deep learning methods for drug response prediction in cancer: predominant and emerging trends}

\usepackage[utf8]{inputenc} 
\usepackage[T1]{fontenc}    
\usepackage{hyperref}       
\usepackage{url}            
\usepackage{booktabs}       
\usepackage{amsfonts}       
\usepackage{nicefrac}       
\usepackage{microtype}      
\usepackage{caption}
\usepackage{subcaption}
\usepackage{comment}
\usepackage{lipsum}
\usepackage[dvipsnames]{xcolor}
\usepackage{multirow}
\usepackage{soul}

\usepackage{graphicx}
\usepackage{afterpage}
\usepackage{array}
\usepackage{booktabs}
\usepackage{amssymb}
\usepackage{amsbsy}
\usepackage{longtable}

\graphicspath{ {./figures/} }

\newcommand{\eg}[0]{\textit{e.g.}}
\newcommand{\ie}[0]{\textit{i.e.}}
\newcommand{\ea}[0]{et al. }

\begin{document}
\maketitle

\begin{abstract}
Cancer claims millions of lives yearly worldwide. While many therapies have been made available in recent years, by in large cancer remains unsolved.
Exploiting computational predictive models to study and treat cancer holds great promise in improving drug development and personalized design of treatment plans, ultimately suppressing tumors, alleviating suffering, and prolonging lives of patients.
A wave of recent papers demonstrates promising results in predicting cancer response to drug treatments while utilizing deep learning methods. These papers investigate diverse data representations, neural network architectures, learning methodologies, and evaluations schemes.
However, deciphering promising predominant and emerging trends is difficult due to the variety of explored methods and lack of standardized framework for comparing drug response prediction models.
To obtain a comprehensive landscape of deep learning methods, we conducted an extensive search and analysis of deep learning models that predict the response to single drug treatments. A total of 60 deep learning-based models have been curated and summary plots were generated. Based on the analysis, observable patterns and prevalence of methods have been revealed. This review allows to better understand the current state of the field and identify major challenges and promising solution paths.

\end{abstract}


\section{Introduction} \label{sec:intro}

Cancer treatment response prediction is a problem of great importance for both clinical and pharmacological research communities. Many believe it will pave the way to devising more efficient treatment protocols for individual patients and provide insights into designing novel drugs that efficiently suppress disease. 
Currently, however, cancer treatment remains extremely challenging, often resulting in inconsistent outcomes. For example, tumor
 heterogeneity contributes to differential treatment responses in patients with the same tumor type \cite{yancovitz_intra-_2012, fisher_cancer_2013}.
Nevertheless, conventional tumor type-dependent anticancer treatments such as chemotherapy often lead to sub-optimal results and substantial side effects, and therefore, are notoriously regarded as one-size-fits-all therapies \cite{collins_towards_2017, hartl_translational_2021}.
%
Alternatively, targeted therapies and certain immunotherapies are prescribed based on known biomarkers, observable within individual patients \cite{prasad_cancer_2018}. 
Cancer biomarkers refer to abnormalities in omics data (genomic, transcriptomic, etc.) which can be predictive of treatment response \cite{collins_towards_2017}. 
Biomarker-driven treatment plans, either standalone or in combination with chemotherapies, are the mainstream of nowadays personalized (or precision) oncology.
%
%
Discovery of biomarkers and their subsequent utilization in clinical settings are attributed to advances in tumor profiling technologies and high-throughput drug screenings \cite{collins_towards_2017, hartl_translational_2021}.

An alternative direction to leverage large-scale screenings and high-dimensional omics data in the cancer research community is to build analytical models designed to predict the response of tumors to drug treatments. Typically, such models use tumor and drug information without explicitly specifying biomarkers \cite{adam_machine_2020}.
These models, often referred to as drug response prediction (DRP) models, can be used to prioritize treatments, explore drug repurposing, and reaffirm existing biomarkers.
Machine learning (ML) and deep learning (DL) algorithms (often referred to as artificial intelligence or AI) are the core methodology in designing DRP models, demonstrating encouraging results in retrospective evaluation analyses with pre-clinical and clinical datasets.

Papers in this field are being published constantly, exploiting learning algorithms for DRP.
To cope with increasing rate of publications, a recent special issue in Briefings in Bioinformatics was dedicated to DRP in cancer models \cite{ballester_artificial_2021}.
Collectively, at least 18 review-like papers have been published since 2020 in an attempt to summarize progress and challenges in the field, as well as provide discussions on promising research directions.
Each paper aims to review the field from a unique perspective but certain topics substantially overlap as shown in Table \ref{tab:drp_review_topics} which lists some of the major topics and associated references.
Common topics include data resources for constructing training and test datasets, prevalent data representations, ML and DL approaches for modeling drug response, and methods for evaluating the predictive performance of models.

\begin{table}
    \caption{Categorization of topics covered in existing review papers related to drug response prediction (DRP).}
\begin{minipage}{16.5cm}
   \centering
   \begin{tabular}{ llc }
       \toprule
       Category & Topic & Review papers \\
       \midrule
       \multirow{1}{*}{Data resources\footnote{Characterization of cancer tumors: resources of cancer omics data; Characterization of drugs: tools for generating drug feature representations; Monotherapy screenings: resources of single-drug response data; Drug combination screenings: resources of drug-combination response data; Complimentary resources relevant to DRP: additional data resources and computational methods relevant to the DRP problem such as drug-target interactions, pathway information, etc.}}
        & Characterization of cancer tumors\hspace{5pt} &
          \cite{de_niz_algorithms_2016, piyawajanusorn_gentle_2021, baptista_deep_2021, kumar_comprehensive_2022} \\&&
          \cite{azuaje_computational_2017, ali_machine_2019, sharifi-noghabi_drug_2021, guvenc_paltun_improving_2021, rafique_machine_2021, zadorozhny_deep_2021, firoozbakht_overview_2022}\vspace{5pt}\\
        & Characterization of drugs\hspace{5pt} &
          \cite{chiu_deep_2020, tanoli_artificial_2021, an_representation_2021} \\&&
          \cite{guvenc_paltun_improving_2021, zadorozhny_deep_2021}\vspace{5pt}\\
        & Monotherapy screenings\hspace{5pt} &
          \cite{piyawajanusorn_gentle_2021, baptista_deep_2021, firoozbakht_overview_2022} \\&&
          \cite{azuaje_computational_2017, ali_machine_2019, chiu_deep_2020, sharifi-noghabi_drug_2021, guvenc_paltun_improving_2021, rafique_machine_2021}\vspace{5pt}\\
        & Drug combination screenings\hspace{5pt} &
          \cite{adam_machine_2020, piyawajanusorn_gentle_2021, baptista_deep_2021, wu_machine_2022, kumar_comprehensive_2022} \\&&
          \cite{chiu_deep_2020}\vspace{5pt}\\
        & Complimentary resources\hspace{5pt} &
          \cite{chiu_deep_2020, adam_machine_2020, wu_single-cell_2020, tanoli_artificial_2021, kumar_comprehensive_2022} \\&&
          \cite{de_niz_algorithms_2016, baptista_deep_2021, wu_machine_2022}\\
       \midrule
       \multirow{1}{*}{Data representations methods\footnote{Representation of cancer tumors: gene expressions, copy-number variations, and other representations of tumors; Representation of drugs: drug molecular descriptors, fingerprints, molecular graph structures, etc.; Representation of treatment response: drug efficacy metrics such as IC50, AUC, sensitive/resistant, and others; Feature selection and extraction: methods for feature selection and dimensionality reduction.}}
        & Representation of cancer tumors\hspace{5pt} &
          \cite{de_niz_algorithms_2016, piyawajanusorn_gentle_2021} \\&&
          \cite{azuaje_computational_2017, chen_survey_2021, firoozbakht_overview_2022}\vspace{5pt}\\
        & Representation of drugs\hspace{5pt} &
          \cite{an_representation_2021, zadorozhny_deep_2021} \\&&
          \cite{chen_survey_2021}\vspace{5pt}\\
        & Representation of treatment response\hspace{5pt} &
          \cite{de_niz_algorithms_2016, sharifi-noghabi_drug_2021, piyawajanusorn_gentle_2021} \\&&
          \cite{chen_survey_2021, firoozbakht_overview_2022}\vspace{5pt}\\
        & Feature selection and extraction\hspace{5pt} &
          \cite{de_niz_algorithms_2016, ali_machine_2019, koras_feature_2020} \\&&
          \cite{adam_machine_2020, rafique_machine_2021}\\
       \midrule
       \multirow{1}{*}{Prediction models\footnote{Models for monotherapy: discussion of models predicting single-drug response; Models for drug-combination therapies: discussion of models predicting drug-combination response.}}
        & DL models for monotherapy\hspace{5pt} &
          \cite{baptista_deep_2021, firoozbakht_overview_2022} \\&&
          \cite{chiu_deep_2020, adam_machine_2020, wu_single-cell_2020, an_representation_2021, sharifi-noghabi_drug_2021, rafique_machine_2021, zadorozhny_deep_2021, chen_how_2021}\vspace{5pt}\\
        & DL models for drug combination therapies\hspace{5pt} &
          \cite{baptista_deep_2021, wu_machine_2022, kumar_comprehensive_2022} \\&&
          \cite{chiu_deep_2020, adam_machine_2020, an_representation_2021, rafique_machine_2021}\vspace{5pt}\\
        & ML models for monotherapy\hspace{5pt} &
          \cite{costello_community_2014, de_niz_algorithms_2016, guvenc_paltun_improving_2021, chen_survey_2021, firoozbakht_overview_2022} \\&&
          \cite{azuaje_computational_2017, ali_machine_2019, adam_machine_2020, an_representation_2021, sharifi-noghabi_drug_2021, rafique_machine_2021}\vspace{5pt}\\
        & ML models for drug combination therapies\hspace{5pt} &
          \cite{menden_community_2019, wu_machine_2022} \\&&
          \cite{adam_machine_2020, an_representation_2021, rafique_machine_2021}\\
       \midrule
       \multirow{1}{*}{Assessment and evaluation\footnote{Experimental validation of prediction models: assessment studies that conduct experiments to assess performance of DRP models; Evaluation methods of model performance: discussion of methods for evaluating performance of DRP models.}}
        & Experimental validation of prediction models\hspace{5pt} &
          \cite{costello_community_2014, menden_community_2019, chen_survey_2021, chen_how_2021} \\&&
          \cite{koras_feature_2020, sharifi-noghabi_drug_2021}\vspace{5pt}\\
        & Evaluation methods of model performance\hspace{5pt} &
          \cite{de_niz_algorithms_2016, sharifi-noghabi_drug_2021, baptista_deep_2021, chen_how_2021, firoozbakht_overview_2022} \\&&
          \cite{azuaje_computational_2017, wu_machine_2022}\\
       \bottomrule
  \end{tabular}
  \label{tab:drp_review_topics}
\end{minipage}
\end{table}

\begin{figure}[h]
    \centering
    \includegraphics[width=0.7\textwidth]{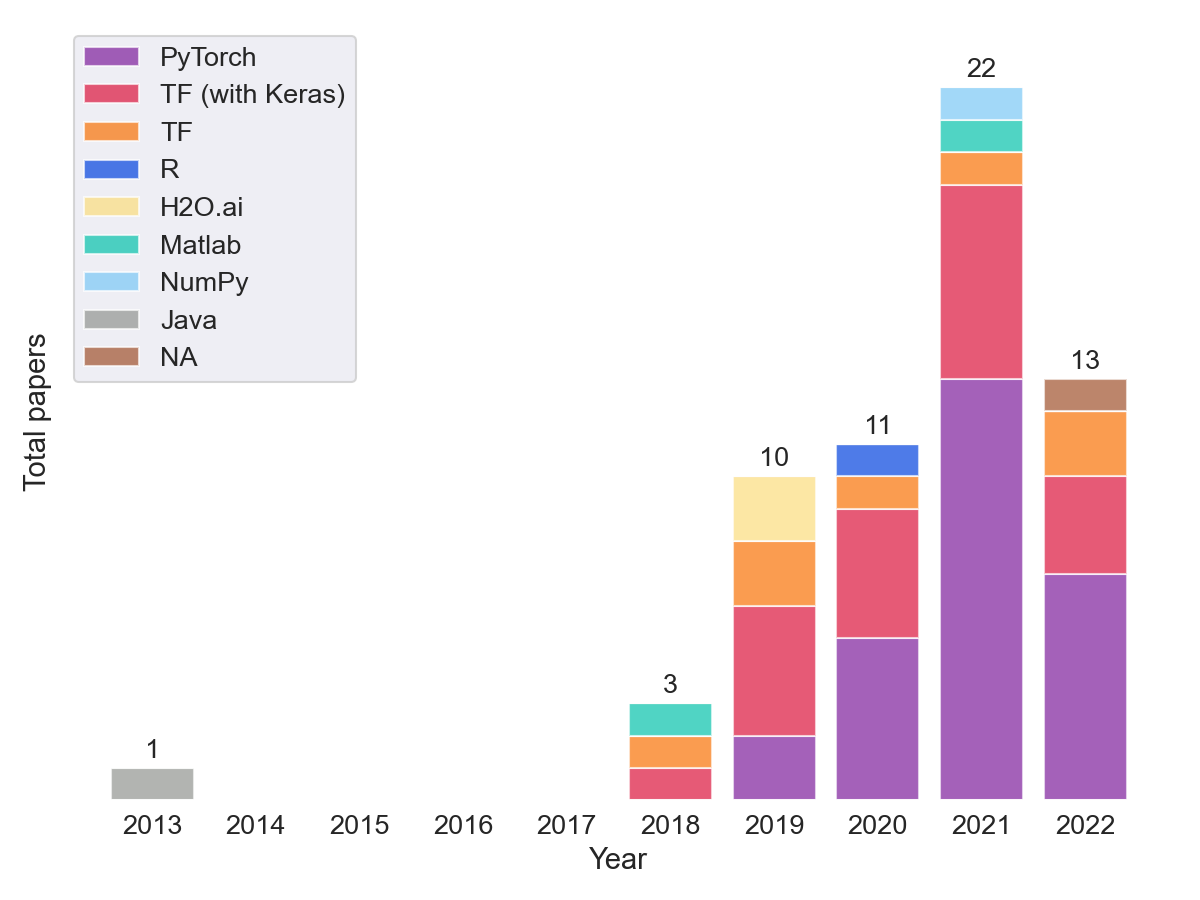}
    \caption{
    Distribution by year of papers that use deep learning methods for drug response prediction (color-coded by deep learning computational frameworks). Data was collected until August of 2022, considering only peer-reviewed publications.
    }
    \label{fig:papers_per_year}
\end{figure}

Following the revival of artificial neural networks (NNs) more than a decade ago \cite{lecun_deep_2015}, DL methods have become a promising research direction across a variety of scientific and engineering disciplines \cite{stevens_ai_2020, rolnick_tackling_2019, cohen_deep_2019, utyamishev2022multiterminal}. 
This trend is also observed in cancer research, including the prediction of tumor response to treatments, as shown in Fig. \ref{fig:papers_per_year}. 
In 2013, Menden \ea demonstrated that a single hidden layer NN predicts drug response with comparable performance to random forest (RF) \cite{menden_machine_2013}.
The authors used genomic and response data from the Genomics of Drug Sensitivity in Cancer (GDSC) project which was published in 2012 \cite{yang_genomics_2013}. 
Despite the accelerated popularity of DL and the availability of substantial screening and omics data, it was not until the emergence of open-source frameworks dedicated for building and training NNs \cite{tensorflow2015-whitepaper, chollet2015keras, NEURIPS2019_9015} that DL has become an integral part of DRP research (note the gap between 2013 and 2018 in Fig \ref{fig:papers_per_year}).
Owing to the abundance of omics and screening data, the availability of DL frameworks and the unmet need for precision oncology, we have been witnessing a growth of scientific publications exploring NN architectures for DRP.

Dozens of DL-based models have been published, exploring diverse feature representations, NN architectures, learning schemes, and evaluation methods.
However, only three out of the many existing reviews explicitly focus on modeling DRP with DL \cite{baptista_deep_2021, kumar_comprehensive_2022, firoozbakht_overview_2022}.
%
These reviews present a categorized summary of methods, citing and discussing a limited number of examples from each category.
As a result, only selective overview of methods is provided and limited aspects of published DRP models are considered, exhibiting a primary limitation of these reviews.
In addition, the scope of existing reviews does not require a comprehensive search for models which could reveal predominant and emerging trends.
%
%
Considering the current rate of publications, the diversity of approaches, and the limitations of existing reviews, a comprehensive review is necessary and timely.

%
This review is the first one to conduct a comprehensive search to include all relevant papers that utilize NNs for DRP.
As of the time of completing this paper, a total of 60 peer-reviewed publications have been identified.
Only models predicting response to single-drug treatments are included in the current review, excluding models that make predictions to combination therapies.
We identified three major components involved in developing DRP models, including data preparation, model development, and performance analysis.
These components have been used to guide the curation of papers with special focus on representation methods of drugs, cancers, and measures of treatment response, DL related methods including NN modules and learning schemes, and methods for evaluating the prediction performance.
This information is summarized in Table S1.
Summary plots have been generated, revealing the prevalence of methods used in these papers.
Observing the prevalence of methods will assist in revealing approaches that have been investigated in multiple studies as well as emerging methods which are rather underexplored for DRP.
We believe this review would serve as a valuable reference for new and experienced researchers in this field.

\begin{figure}[h]
    \centering
    \includegraphics[width=0.95\textwidth]{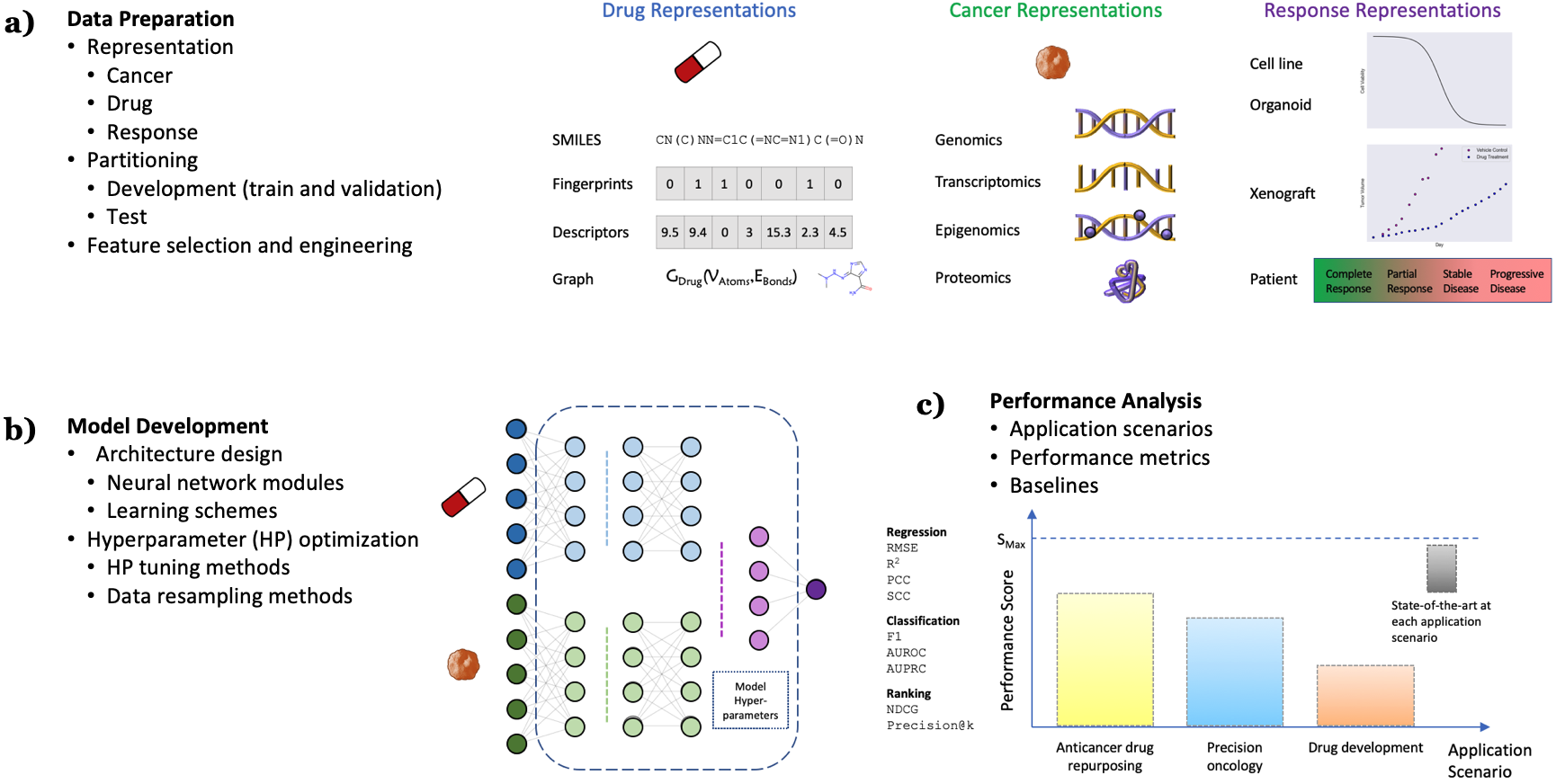}
    \caption{
    General components of a drug response prediction (DRP) workflow. a) Data preparation: requires to generate representations of features and treatment response, partition the dataset into development set (for training and hyperparameter (HP) tuning) and test set (for performance analysis), and any additional preprocessing such feature selection/engineering. b) Model development: the process of generating a deep learning model which involves the design of a neural network (NN) architecture (choice of NN modules and learning schemes) and model training including HP optimization. c) Performance analysis: assessment of prediction generalization and other metrics allowing to evaluate the utility of the DRP model for different applications in oncology such as personalized recommendation of treatments, drug repurposing, and drug development. The performance is benchmarked against one or more baseline models which should ultimately be chosen from available state-of-the-art models for the investigated application.
    }
    \label{fig:drp_schematic}
\end{figure}

Section \ref{sec:DRP_workflow} formulates the DRP as a DL problem. Sections \ref{sec:rsp_rep}-\ref{sec:drugs_rep} provide a detailed review of the drug, cancer, and response representations used in papers listed in Table S1. Existing design choices are summarized in Sections \ref{sec:dl_modules} and \ref{sec:dl_learning_schemes} in terms of fundamental NN building blocks and learning schemes, respectively. Section \ref{sec:eval_methods} compiles existing approaches for evaluating model performance. In Section \ref{sec:discussion}, we discuss the current state of DRP field, determine primary challenges, and propose further directions.

\section{Deep learning-based drug response prediction workflow} \label{sec:DRP_workflow}

A DRP model can be represented by $r = f(d, c)$, where $f$ is the analytical model designed to predict the response $r$ of cancer $c$ to the treatment by drug $d$.
The function $f$ is implemented with a NN architecture in which the weights are learned through backpropagation.
This formulation is for pan-cancer and multi-drug\footnote{sometimes referred as pan-drug} prediction model where both cancer and drug representations are needed to predict response.
A special case is drug-specific models designed to make predictions for a drug or drug family (\eg, drugs with the same mechanism of action (MoA)) \cite{sharifi-noghabi_moli_2019}. These models learn from cancer features only and can be formulated as $r=f_D(c)$. 

The general workflow for developing DRP models is not much different from developing supervised models for other applications (Fig \ref{fig:drp_schematic}).
The challenges come in the specific details typical to predicting treatment response.
The process can be divided into three components: 1) data preparation, 2) model development, and 3) performance analysis \cite{azuaje_computational_2017, baptista_deep_2021}.
The choice of methods associated with each one of these components can have a significant impact on the overall workflow complexity and the potential application of the final model.

\textbf{Data preparation.}
Data preparation is generally the initial step in designing a prediction model, requiring expertise in bioinformatics and statistical methods. During this step, heterogeneous data types are aggregated from multiple data sources, preprocessed, split into training and test sets, and structured to conform to an API of a DL framework.
The generated drug response dataset with $N$ samples, denoted by $S=\{d, c, r\}_{i=1}^N$, includes representations for drug ($d$), cancer ($c$), and response ($r$).
Prediction generalization is expected to improve with a larger number of training samples as demonstrated with multiple cell line datasets \cite{partin_learning_2021, xia_cross-study_2022, prasse_pre-training_2022}, normally creating preference for larger datasets when developing DL models.
Recent research focusing on data-centric approaches suggests that efficient data representations and proper choice of a training set are at least as important as the dataset size for improving predictions and further emphasize the importance of the data preprocessing step \cite{mazumder_dataperf_2022}.

\textbf{Model development.}
Model development refers to NN architecture design and optimization of model hyperparameters (HPs).
To design NNs, developers often resort to common heuristics which rely on intuition, experimentation, and adoption of architectures from related fields.
This process involves choosing the basic NN modules, the architecture, and learning schemes.
Diversity of data representations for cancers \cite{piyawajanusorn_gentle_2021} and drugs \cite{an_representation_2021} and potential utilization of DRP models in several pre-clinical and clinical settings have led researchers to explore a wide range of DL methods.

\textbf{Performance analysis.}
A desirable outcome of a model development workflow is a robust model that produces accurate predictions across cancers and drugs as evaluated by appropriate performance metrics.
DRP models can be used in various scenarios such as personalized recommendation of treatments, exploration of drug repurposing, and assisting in development of new drugs.
Therefore, both performance metrics and appropriate evaluation schemes (\eg, design of training and test sets) are critical for proper evaluation of prediction performance.
The abundance of DRP papers in recent years (Fig. \ref{fig:papers_per_year}) and lack of benchmark datasets \cite{baptista_deep_2021}, strongly suggest that a rigorous assessment of model performance is required where state-of-the-art baseline models serve as a point of reference.

\section{Representations of treatment response} \label{sec:rsp_rep}

\begin{figure}[h]
    \centering
    \includegraphics[width=0.7\textwidth]{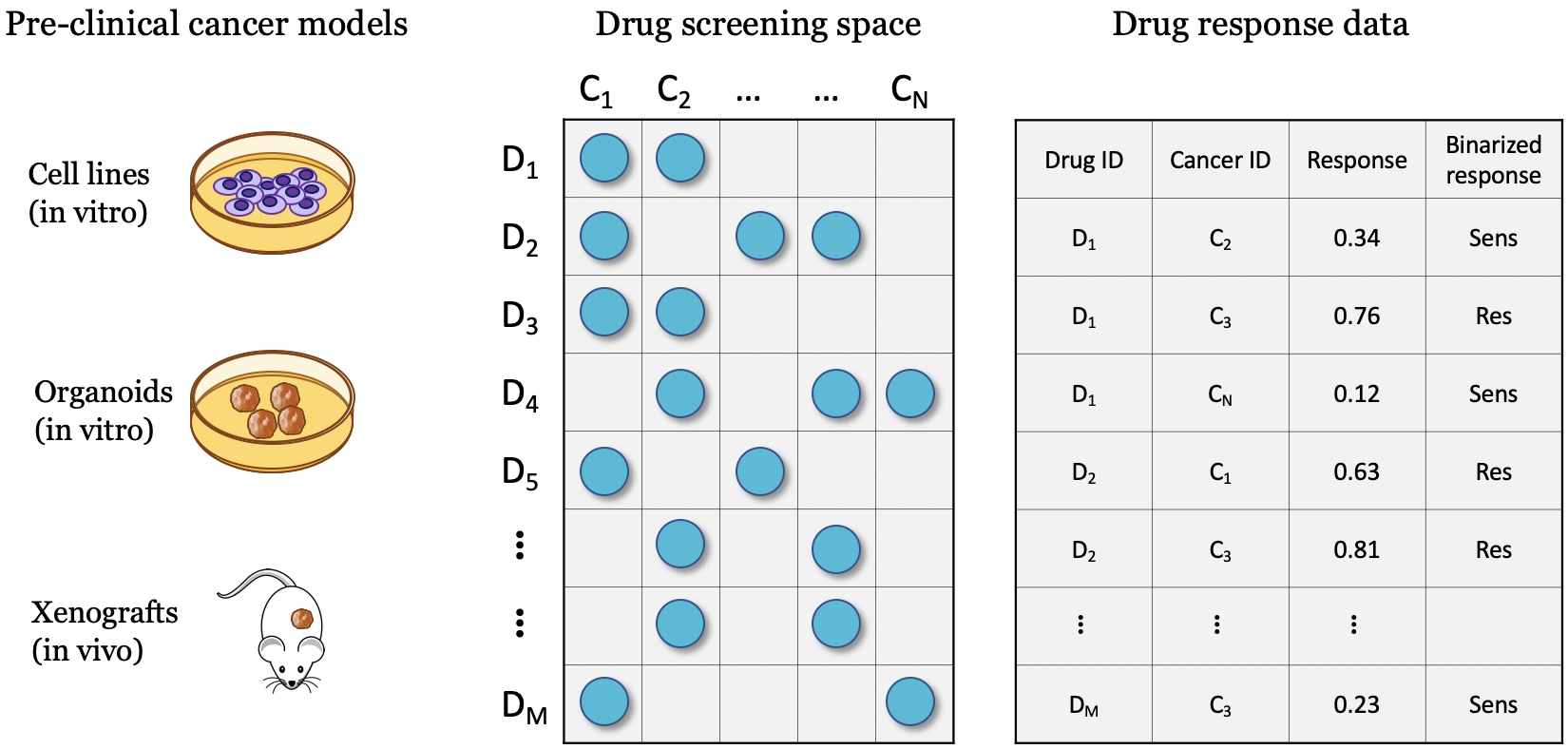}
    \caption{
    Drug screening experiments are performed with various cancer models such as cell lines, organoids, and xenografts, where cancer samples are screened against a library of drug compounds. The screening data is transformed into a drug response dataset that can used for developing drug response prediction models, including regression, classification, and ranking.
    }
    \label{fig:screening_data}
\end{figure}

%
Drug screening platforms enable testing the sensitivity of cancer samples in controlled lab environments, ultimately producing data for predictive modeling (Fig. \ref{fig:screening_data}).
A major objective is a discovery of potential anticancer treatments through the screening of compound libraries against diverse cancer panels. 
Systematic drug screening platforms have been established for in vitro cancer models such as cell lines \cite{barretina_cancer_2012, iorio_landscape_2016, seashore-ludlow_harnessing_2015} and organoids \cite{larsen_pan-cancer_2021}, and in vivo models such as patient-derived xenografts (PDXs) \cite{gao_high-throughput_2015}.
An overview of preclinical cancer models and the corresponding methods for sensitivity profiling is available in a recent review article \cite{piyawajanusorn_gentle_2021}.
Drug sensitivity data can be transformed into continuous or categorical variables, representing the treatment response.
Accordingly, prediction models have been designed to solve regression, classification, and ranking problems.
This section reviews methods for representing treatment response and the corresponding prediction tasks.

\subsection{Cell lines}
Cell line studies constitute the most abundant resource of response data.
High-throughput drug screenings with cell lines is performed with a compounds library, where each cell-drug combination is screened at multiple drug concentrations. The response of in vitro cells at each concentration is assessed via a cell viability assay which quantifies the surviving (viable) cells after treatment versus the untreated control cells. Performing the experiments over a range of concentrations results in a vector of non-negative continuous dose-response values for each cell-drug pair.
Those data points are summarized via a dose-response curve obtained by fitting a four parameters logistic Hill equation.

\subsubsection{Continuous measures of response}
\begin{figure}[h]
    \centering
    \includegraphics[width=0.7\textwidth]{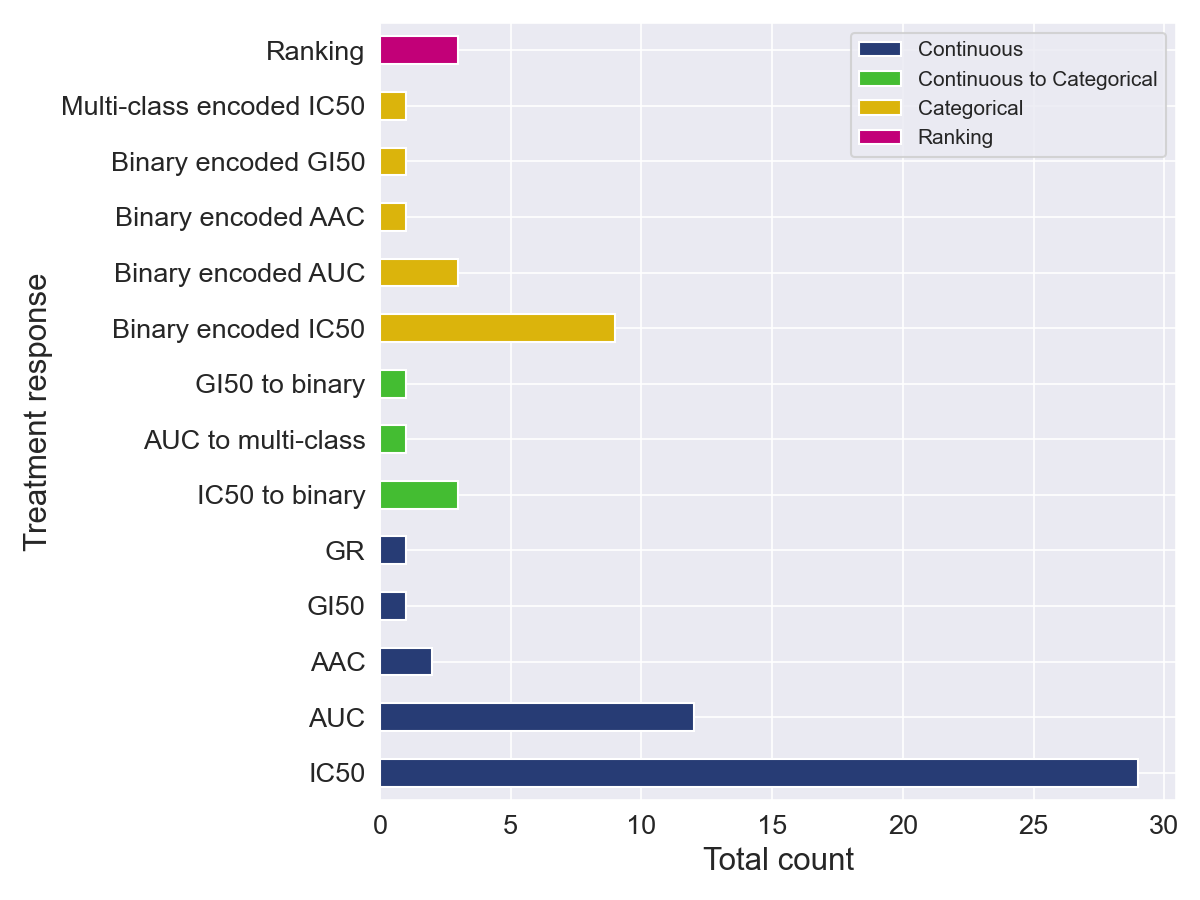}
    \caption{
    Cell-line drug response data is usually represented with continuous or categorical values. Drug response prediction (DRP) models use the different drug response representations to train regression, classification and ranking models. The histogram illustrates the prevalence of the difference representations and learning tasks. Certain papers exploit several representation of response and learning tasks, and therefore, these papers contribute more than one item to the histogram.
    The label \textit{categorical} means that the continuous response was first categorized and then a DRP classifier was trained, while \textit{continuous to categorical} means that a DRP regressor was trained and then the predicted response was categorized (in both cases classification metrics were used for performance analysis).
    }
    \label{fig:rsp_cell}
\end{figure}

The cell line dose-dependent responses lack a direct translation into the space of in vivo cancer models.
A common approach is to extract from the dose-response curve a single-value summary statistic which represents the response for a cell-drug pair.
Several methods exist allowing to calculate continuous response values which serve as the prediction target with supervised regression models.
The two most common metrics are IC50 and AUC, with IC50 being substantially more prevalent (Fig. \ref{fig:rsp_cell}).

Observing such predominance of IC50, an immediate question is whether IC50 exhibits substantial benefits over other measures. The half-maximal inhibitory concentration, \ie, IC50, is the concentration at which the drug reaches half of its maximal inhibition power on the fitted dose-response curve.
Alternatively, measures such as AUC (area under the dose-response curve), AAC (area above the dose-response curve or activity area), and DSS (drug sensitivity score), are obtained by aggregating the cell viability values across a range of concentrations of the dose-response curve, providing what is considered a more global measure of response.
Arguments in favor of these global measures are available, discussing the benefits of AUC \cite{jang_systematic_2013, xia_cross-study_2022} and DSS \cite{yadav_quantitative_2014} as opposed to IC50.
In addition, empirical analyses suggest better prediction generalization in the case of using AAC values as opposed to the alternative of using IC50 \cite{sharifi-noghabi_drug_2021}.

\subsubsection{Categorical measures of response}
Despite the prevalence of IC50 and AUC, there are arguments suggesting that continuous measures of response (blue in Fig \ref{fig:rsp_cell}) lack a straightforward interpretation in the context of decision-making purposes.
Alternatively, a categorical output representing a discrete level of response such as \textit{sensitive} versus \textit{resistant} is more comprehensible for humans, and thus, better supports actionable outcomes.
Two primary approaches were used to produce categorical responses with DRP models.


In the more common approach, continuous responses were first categorized usually using one of the available methods such as waterfall algorithm \cite{barretina_cancer_2012, haibe-kains_inconsistency_2013}, LOBICO \cite{knijnenburg_logic_2016, iorio_landscape_2016}, or a histogram-based method \cite{staunton_chemosensitivity_2001, dong_anticancer_2015}.
DL classifiers were trained on the transformed values to predict class probability, which is subsequently translated into a discrete response label (orange in Fig. \ref{fig:rsp_cell}).
The predicted probability can also be utilized as a quantitative measure of prediction uncertainty, an essential aspect in decision making. 
Most models were trained to predict a binary response, representing that cancer is either sensitive or resistant to treatment.
A multi-class classifier with three classes corresponding to low, intermediate, and high responsiveness, was also explored \cite{joo_deep_2019}.
The second approach is to train regression models to predict one of the continuous responses and then categorize it into two or more classes (green in Fig. \ref{fig:rsp_cell}). Only a few papers explored this approach, including binary \cite{sakellaropoulos_deep_2019, daoud_q-rank_2020, malik_deep_2021, jiang_deeptta_2022} and multi-class labels \cite{zhao_computational_2019}.
This approach naturally allows assessing model performance using both regression and classification metrics, possibly offering a more robust generalization analysis.


\subsubsection{Ranking}



In addition to producing continuous or categorical predictions, models can be trained to learn a ranking function with drug response data.
%
%
In the context of clinical precision oncology, the clinician might be interested in obtaining a ranked list of the top-k drugs that are likely to be most beneficial for the patient.
In our search, we obtained three ranking models all of which targeting personalized treatment recommendation.
By framing the problem as a ranking task, the authors proposed models that learn to produce a ranked list of drugs per cell-line that are most likely to inhibit cell viability.
Prasse et al. transformed IC50 into drug relevance scores and derived a differentiable optimization criteria to solve the ranking problem which can be combined with different NN architectures \cite{prasse_matching_2022}.
By combining their ranking learning method with the PaccMann architecture \cite{manica_toward_2019} and an fully-connected NN (FC-NN), they significantly improved the ranking performance as compared to baseline models.
SRDFM useds a deep factorization machine (DeepFM) with pairwise ranking approach which generates relative rankings of drugs rather than exact relevance scores \cite{su_srdfm_2022}.
PPORank generates drug rankings using deep reinforcement learning which enables to sequentially improve the model as more data becomes available.
SRDFM and PPORank rankings outperform non-DL models.


\subsection{Patient derived xenografts and patient tumors}

Whereas cell line data serve as the primary resource for training DL models, several papers proposed methods for predicting response in PDX and patient tumors.
PDX is a contemporary cancer model that was developed to better emulate human cancer in medium-scale drug screenings, providing a controlled environment for studying the disease and systematically testing treatments in pre-clinical settings.
While systematic drug screenings with patients is practically impossible \cite{adam_machine_2020}, public data containing treatment response in individual patients are available.

Tumor response in humans is obtained via a standardized evaluation framework called Response Evaluation Criteria in Solid Tumors (RECIST) \cite{schwartz_recist_2016}. RECIST involves non-invasive imaging followed by evaluation of change in tumor size. Four categories used for grading tumor change include Complete Response (CR), Partial Response (PR), Progressive Disease (PD) and Stable Disease (SD). 
Tumor change in PDXs is evaluated by monitoring tumor volume over time using more traditional methods (\eg, calipers) as opposed to using RECIST, primarily due to cost.
The main data resource for modeling drug response with PDXs, provides continuous response values calculated based on tumor volume \cite{gao_high-throughput_2015}.
To create a counterpart to RECIST, they also set cutoff criteria allowing to transform the continuous response values into the four categories. 

Because discrete response labels are available in patient and PDX datasets, DRP models are primarily designed as classifiers where the four-label RECIST categories are transformed into sensitive (CR and PR) and resistant (PD and SD) labels.
Small sample size, however, is a major challenge in developing DL models with these datasets. This issue is generally addressed by incorporating abundant cell line data and transfer learning schemes to improve predictions in PDX and patients (discussed in Section \ref{sec:transfer_learning}).
\section{Cancer representations} \label{sec:tissues_rep}

As personalized oncology treatments rely on omics biomarkers, progress in high-throughput tissue profiling plays an important role in existing and future cancer therapies.
Pharmaco-omic studies conduct multiomic profiling of cancer models and drug screening experiments \cite{piyawajanusorn_gentle_2021}.
Advances in sequencing technologies allowed to substantially scale the experimental studies by increasing the throughput rate and reducing the cost of profiling. The algorithms for processing the raw data evolve as well, thereby providing more reliable representations of the underlying cancer biology and allowing projects to update their repositories with refined versions of existing data while utilizing improved processing techniques (\eg, DepMap portal \cite{depmap}).

\begin{figure}[h]
    \centering
    \includegraphics[width=0.5\textwidth]{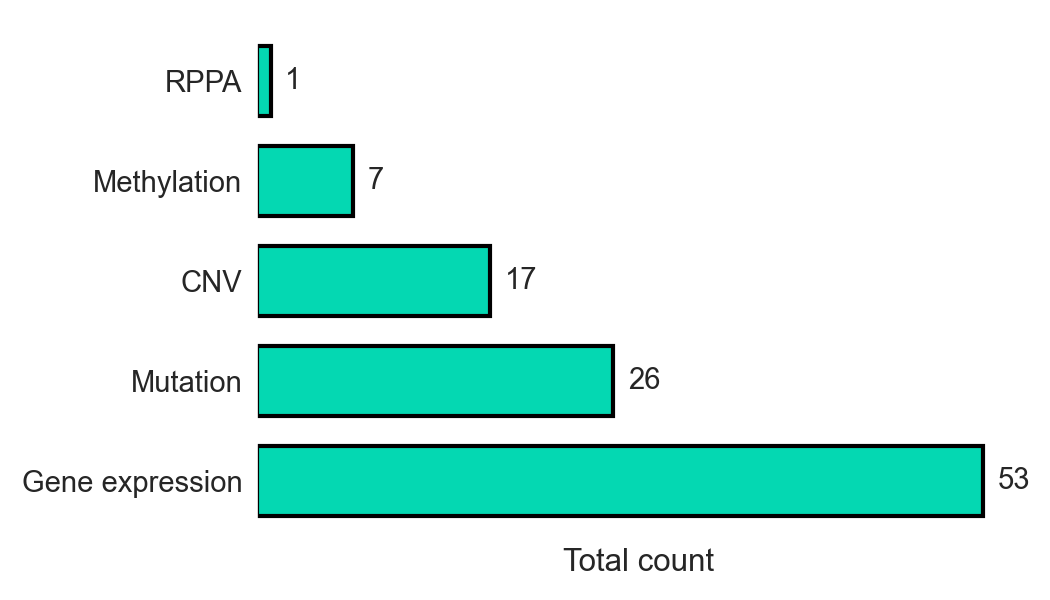}
    \caption{
    Prevalence of omic representations used as features in deep learning-based drug response prediction models. CNV: copy number variations, RPPA: reverse phase protein arrays.
    }
    \label{fig:fea_cancer_all}
\end{figure}

Dedicated bioinformatics pipelines for transforming raw data depend on the omic type and the profiling technology where common steps include alignment, quantification, normalization, and quality control.
Profiling at each omic level produces high-dimensional representation of cancer that can be used as features in the downstream modeling of drug response.
Existing prediction models utilize features that were obtained primarily at four omic levels including genomic (mutation, copy-number variation (CNV)), transcriptomic (gene expression microarrays, RNA-Seq), epigenomic (methylation), and proteomic (Reverse Phase Protein Arrays (RPPA)) \cite{piyawajanusorn_gentle_2021}.
Fig. \ref{fig:fea_cancer_all} shows the prevalence of these representations in DRP models.
Many models use these omic features directly as inputs to NNs while others optimize representations via additional preprocessing with the goal of increasing predictive power \cite{bazgir_representation_2020, zhu_converting_2021}.

In 2014, the NCI-DREAM challenge was a community effort that assessed performance of various classical ML models in predicting drug response in breast cancer cell-lines \cite{costello_community_2014}. It was reported that for most models participated in the challenge, gene expression microarrays provided greater predictive power than other data types. Results of this challenge seem to set the tone for future research in this field as gene expression was used in approximately 90\% of the models either alone or in combination with other feature types, including mutation, CNV, methylation, and RPPA. 
Mutation and CNV are also common but rarely used without gene expression (with only 12\% of such models). 

\begin{figure}[h]
    \centering
    \includegraphics[width=0.7\textwidth]{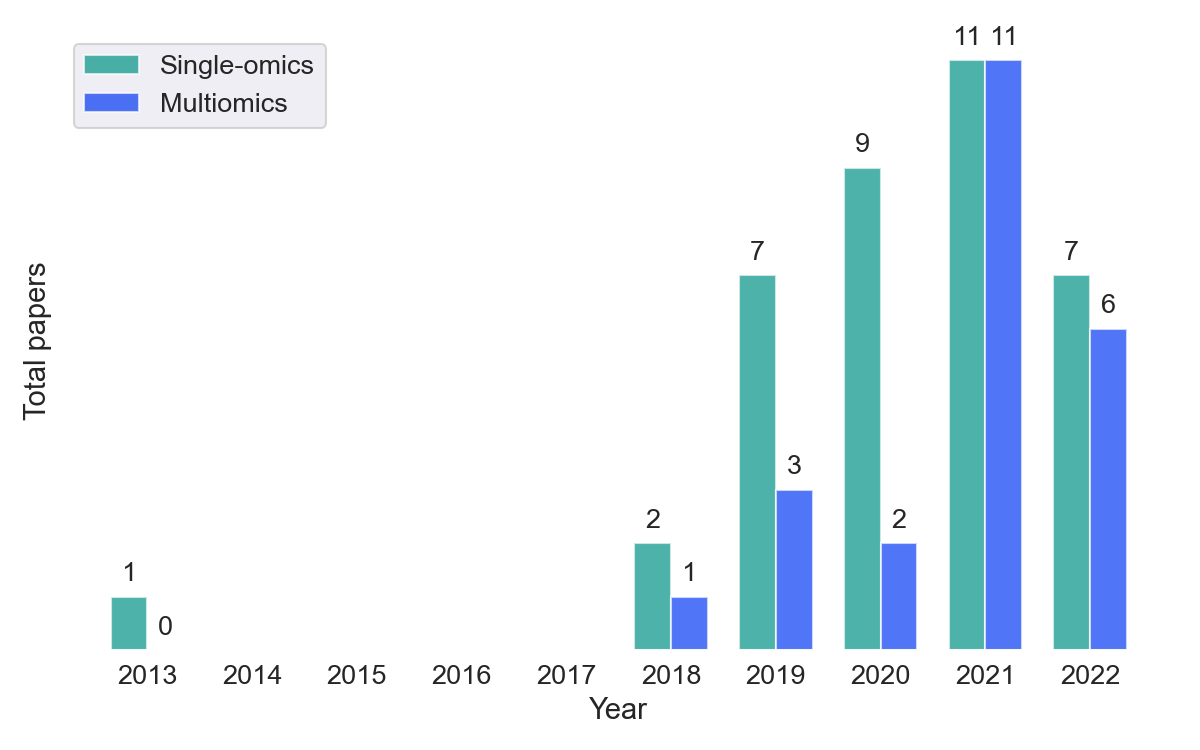}
    \caption{
    Multiomic and single-omic data were explored as features for representing cancer samples in drug response prediction models. Data was collected until August of 2022, considering only peer-reviewed publications.
    }
    \label{fig:multiomics_and_singleomics}
\end{figure}

Learning from multiple omic types have shown to improve generalization, and as consequence, integration of multiomics have been a recent trend (Fig. \ref{fig:multiomics_and_singleomics}).
A straightforward approach is to concatenate the multiomic profiles to form a feature vector and pass it as an input to a NN model, a method commonly referred to as \textit{early integration}.
It has been shown, however, that \textit{late integration} significantly improves predictions where the different omic profiles are passed through separate subnetworks before the integration \cite{sharifi-noghabi_moli_2019, zhu_ensemble_2020, partin_learning_2021}.
Yet, caveats related to data availability hinder the pertinence of multiomics as opposed to using single-omics.
Not having all omics available for all samples is a common occurrence in pharmaco-omic projects.
A common approach to address this issue is filtering the dataset to retain a subset of cancer samples that contain all the required omic types as well as drug response data.
The multiomics data is closely related to multimodal learning which refers to prediction models that learn jointly from multiple data modalities (\ie, representations). It was demonstrated that leveraging multimodal data generally improves predictions.

\section{Representations of drug compounds} \label{sec:drugs_rep}


Contemporary cancer treatment often involves administering therapeutic drugs to patients to inhibit or stop growth of cancer cells, destroy tumors, or boost cancer-related immune system. 
Drug molecules are 3-D chemical structures consisting of atoms and bonds with complex atomic interactions.
Developing numerical representation of molecules is an active research area in several related disciplines, including in silico design and discovery of anticancer drugs.

Ultimately, drug representations should be able to encode essential physical and chemical properties of molecules in compact formats.
Alternatively, when using these representations as drug features for DRP, the model should be able to properly ingest these features and extract information predictive of treatment response.
Therefore, the inherent information encoded with numerical drug representations and the model capability to learn from that specific representation are closely related and mutually important for producing efficient DRP models.

Neural networks provide substantially more flexibility than classical ML in learning from unstructured data such as strings, images, and graphs. This flexibility facilitates a noticeable trend of exploring various drug representations as features in DL-based DRP models \cite{an_representation_2021}. 
Primary types of drug representations include SMILES, fingerprints, descriptors, and graph-based structures (Fig \ref{fig:fea_drug_all}).

\subsection{SMILES}

Perhaps the most common format for querying, handling and storing molecules when working with modern chemoinformatics software tools and databases is a linear notation format called SMILES (simplified molecular-input line-entry system). With this data structure, each molecule is represented with a string of symbolic characters generated by a graph traversal algorithm \cite{david_molecular_2020, an_representation_2021}.
Although the use of SMILES in DRP is quite limited as compared to other applications (\eg, molecule property prediction, QSAR), this format provides several benefits.

Following common text preprocessing steps (\eg, tokenization, one-hot encoding), SMILES can be naturally used with common sequence-aware modules such as RNN \cite{oskooei_paccmann_2018, manica_toward_2019, prasse_matching_2022} and 1-D CNN \cite{liu_anti-cancer_2018}.
Moreover, each molecule can be represented with different strings depending on the initial conditions when applying the SMILES generating algorithm (\ie, starting node of graph traversal).
The resulting strings are referred to as randomized or enumerated SMILES and have been reported to improve generalization when utilized for drug augmentation \cite{bjerrum_smiles_2017, arus-pous_randomized_2019}.
SMILES are less prevalent in DRP models but remain a popular notation for describing molecules because it often serves as an intermediate step for generating other representations such as fingerprints, descriptors, and graph structures.

\subsection{Descriptors and fingerprints}

\begin{figure}[h]
    \centering
    \includegraphics[width=0.5\textwidth]{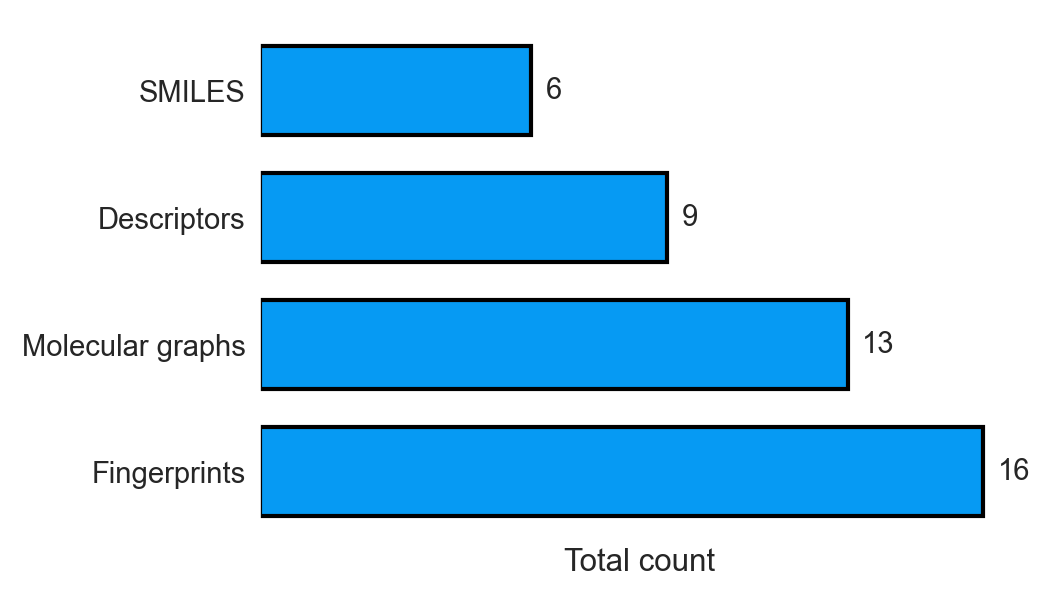}
    \caption{Prevalence of drug feature representations.}
    \label{fig:fea_drug_all}
\end{figure}

Fingerprints (FPs) and descriptors are the two most common feature types for representing drugs in DRP papers (Fig. \ref{fig:fea_drug_all}).
Unlike with SMILES where the string length varies for different molecules, we can specify the same number of features for all drugs in a dataset with either FPs or descriptors. 
The consistent feature dimensionality across drugs 
makes descriptors and FPs easy to use with NNs and classical ML.
With FPs, a drug is a binary vector where each value encodes presence or absence of a molecular substructure with a common vector size of 512, 1024, or 2048.
Multiple algorithms for generating FPs are available implemented by several chemoinformatics packages.
For example, Morgan FPs refers to Extended Connectivity Fingerprints (ECFPs) generated via the Morgan algorithm. ECFP is a class of circular FPs where atom neighborhoods are numerically encoded with binary values. The open-source package RDKit provides an API for generating these FPs which are often used for DRP \cite{rdkit}.
Descriptors is a vector of continuous and discrete values representing various physical and chemical properties, usually containing hundreds or a few thousands of variables. 
Both open-source and proprietary software tools are available for generating descriptors where PaDEL \cite{yap_padel-descriptor_2010}, Mordred \cite{moriwaki_mordred_2018}, and Dragon \cite{dragon} being particularly used for DRP.
A systematic assessment of various features and hyperparameters suggests that DRP with DL exhibits no significant difference when utilizing ECFP, Mordred, or Dragon representations as drug features \cite{clyde_systematic_2020}.



\subsection{Graph structures}


Graphs are powerful representations where complex systems can be represented using nodes and edges.
Recent advancements in GNNs have opened promising research directions allowing efficient learning of predictive representations from graph data \cite{kipf_semi-supervised_2017, velickovic_graph_2018, hamilton_inductive_2018}, and contributing to the widespread interest in GNN among AI researchers and application domain experts \cite{zitnik_modeling_2018, utyamishev2022knowledge}. 
Graph-based representations have emerged as new trend in computational drug development and discovery \cite{sun_graph_2020, gaudelet_utilizing_2021}.
In graph notation, each molecule is described with a unique graph of nodes and edges, where each atom (\ie, node) and each bond (\ie, edge) can be represented with multiple features characterizing physical and chemical properties.
Resources and examples for constructing graphs are well documented in popular chemoinformatics toolkits as well as software packages dedicated for GNNs \cite{Ramsundar-et-al-2019, grattarola_graph_2020, li_dgl-lifesci_2021}.

By borrowing methodologies from related fields such as drug property prediction and drug design, developers of DRP models exploit graph molecular structures for representing drugs combined with GNN-based architectures.
We observe concordant results among those papers investigating molecular graphs with GNNs, reporting superior performance of their models compared to baselines that use non-graph representations.
Based on these recent results and the assumption that graphical description is arguably a more natural way to represent drugs as compared to the aforementioned alternatives, molecular graphs combined with GNNs should be a default modeling choice.
However, only few papers actually report results with rigorous ablation analysis with extensive HP tuning comparing molecular graphs with other representations such as SMILES, FPs, and descriptors for the application of DRP.




\section{Deep learning methods for drug response prediction} \label{sec:dl_methods}

\begin{figure}[h]
    \centering
    \includegraphics[width=1.0\textwidth]{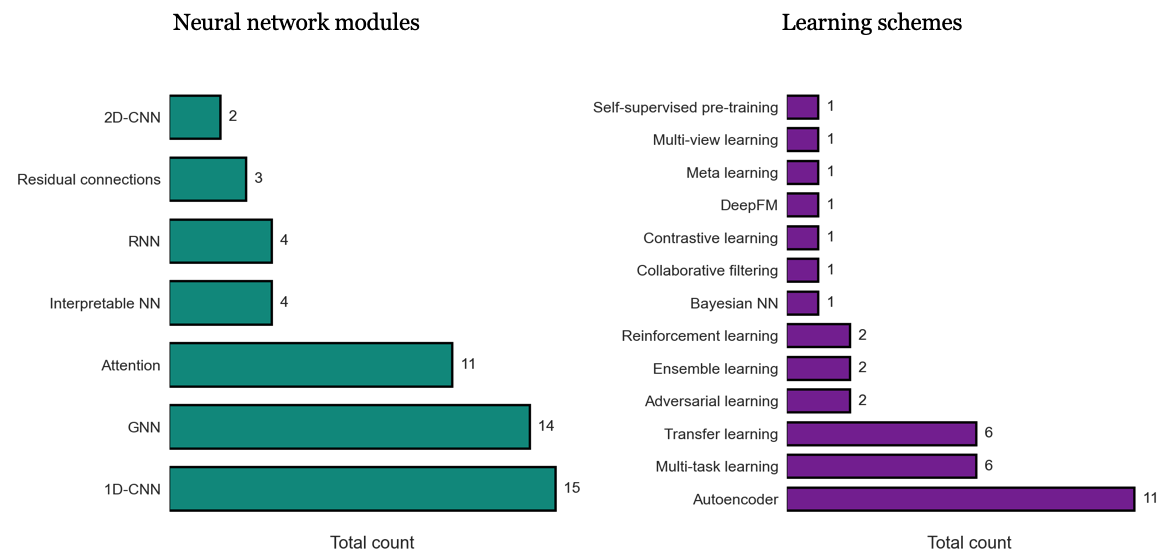}
    \caption{
    Prevalence of deep learning methods, including: a) neural network (NN) modules, and b) learning schemes. 1D-CNN and 2D-CNN: one- and two-dimensional convolutional NN, GNN: graph NN, DeepFM: deep factorization machine.
    }
    \label{fig:DL_methods}
\end{figure}

We review existing approaches for modeling DRP with DL in terms of two perspectives: 1) NN modules that are used as building blocks for constructing DRP architectures, 2) learning schemes that are used to train models with the goal to improve prediction generalization.
The diversity of methods is highlighted in Fig. \ref{fig:DL_methods}.

\subsection{Neural network modules} \label{sec:dl_modules}
Constructing a NN architecture requires choosing appropriate components, their properties, and the way these components are organized and connected (\ie, network topology). 
These components are often available as modules in DL frameworks and range from simple dense layers to more advanced structures such as attention. Certain structures, however, are not generally available as is and should be constructed manually using the available modules (\eg, residual connections).
While there is an absence of rigorous methodologies for designing network topology, the choice of certain modules can be driven by the characteristics of input data.

\subsubsection{Dense layers} \label{sec:dense}
Menden et al. is a pioneering work where cell line and drug features were used to train an FC-NN with a single hidden layer to predict IC50 \cite{menden_machine_2013}.
Recently, FC-NN models with multiple dense hidden layers have been proposed to predict response in cell lines \cite{zhao_computational_2019} and humans \cite{sakellaropoulos_deep_2019}.
Several other models with only dense layers were explored combining advanced NN attributes.
MOLI utilizes triplet loss function and late integration of multiomic inputs to generalize across cancer models by training with cell lines and predicting in PDXs and patients \cite{sharifi-noghabi_moli_2019}.
DrugOrchestra is multi-task model that jointly learns to predict drug response, drug target, and side effects \cite{jiang_drugorchestra_2020}.
RefDNN explores data preprocessing techniques to generate more predictive cell line and drug representations \cite{choi_refdnn_2020}.
For cell line representation, multiple ElasticNets produce a vector representing drug resistance of a cell line to a set of reference drugs, while for drug representation, structure similarity profile is computed for each input drug with respect to the set reference drugs.
Another model with late integration of multiomic data is proposed for predicting drug response and survival outcomes which also uses neighborhood component analysis for feature selection of multiomic data \cite{malik_deep_2021}.
PathDSP utilizes multi-modal data preprocessed via pathway enrichment analysis and integrated into an FC-NN for DRP \cite{tang_explainable_2021}.
All the aforementioned models demonstrate promising prediction performance while utilizing dense layers only in their architectures.

\subsubsection{Convolutional layers} \label{sec:cnn}
Due to state-of-the-art performance in various applications \cite{kiranyaz_1d_2021} and fewer learning parameters required as compared to dense layers, 1-D CNNs are a popular choice in DRP models particularly for processing omics data.
DRP models utilizing CNNs have been published every year since 2018.

DeepIC50 is a multi-class classifier exploiting an early integration of cell line mutations and drug features (descriptors and FPs) passed to a three-layer CNN and followed by an FC-NN prediction module \cite{joo_deep_2019}.
tCNNs exploits late integration of two parallel CNN subnetworks with three layers each \cite{liu_improving_2019}. Binary features of cell lines (mutations and CNV) and drugs (one-hot encoded SMILES) are propagated through the respective subnetworks, concatenated, and passed through an dense layer for the prediction of IC50.
GraphDRP is very similar to tCNNs in terms of CNN utilization, with the primary difference being the utility of molecular graphs and GNN layers for learning drug representations. 
CDRScan is an ensemble of five different architectures with varying design choices (late and early integration, shallow and deep NNs, with and without dense layers) which all contain CNNs and learn to predict IC50 from mutation and FP data \cite{chang_cancer_2018}.

DeepCDR takes the design further by exploiting late integration of drug and multiomic features, including genomic mutation, gene expression, and DNA methylation, where only the mutations are passed through a CNN layers \cite{liu_deepcdr_2020}.
Following DeepCDR, two similar models were published, GraTransDRP \cite{chu_graph_2022} and GraOmicDRP \cite{nguyen_integrating_2022} which both use late integration of multiomics and molecular graphs for drug representation. While DeepCDR uses CNNs only for genomic features, all three subnetworks for multiomic inputs consist of CNNs in GraTransDRP and GraOmicDRP.
SWNet is another model utilizing late integration of cell lines and drugs with three CNN layers in the cell-line subnetwork and one such layer in the prediction subnetwork \cite{zuo_swnet_2021}.

We may gather that 1-D CNNs carry substantial benefits in the context of DRP considering their prevalence (Fig. \ref{fig:DL_methods}).
Based on our search, however, only two papers explicitly report their empirical findings assessing CNNs versus alternative learning modules. Interestingly, both papers suggest that CNNs underperform dense layers in their respective architectures.
Zhao et al. reports that in a twelve-layer model predicting response from gene expression, dense layers outperform CNNs or RNNs \cite{zhao_computational_2019}.
Manica et al. present extensive analyses exploring various architectures for encoding drug representations directly from SMILES, including CNNs, bidirectional RNNs (bRNNs), and various attention modules \cite{manica_toward_2019}.
Their results suggest that a combination of convolutional and attentions modules produces the most predictive model while a CNN-only encoding subnetwork performs the worst as compared to the explored variants.

With the objective to leverage 2-D CNNs, two algorithms were recently published for converting tabular data into images and utilizing CNNs for predicting drug response. Both algorithms, REFINED \cite{bazgir_representation_2020} and IGTD \cite{zhu_converting_2021}, convert gene expressions and drug descriptors into images as a preprocessing step, and then learn to predict drug response with late integration of 2-D CNN subnetworks. The papers report superior performance of proposed algorithms as compared to various baseline models.


\subsubsection{Attention mechanism} \label{sec:attention}
Presenting PaccMann \cite{oskooei_paccmann_2018}, Oskooei et al. were the first to report the use of attention-based NN for DRP, exploiting late feature integration with attention mechanisms incorporated both in the cell line and drug subnetworks. On the cell line path, gene expressions are encoded with self-attention producing a gene attention (GA) vector. On drug path, SMILES embeddings are 
combined with GA via contextual-attention where the GA vector serves as the context. 
This design which was further described in \cite{manica_toward_2019} improves generalization as compared to baselines as well as facilitates interpretability via attention weights.



Attention-based designs have been further explored in recent years.
CADRE is a collaborative filtering model with contextual-attention and pre-trained gene embeddings (gene2vec), designed to recommend treatments based on cell line gene expressions \cite{tao_predicting_2020}.
Ablation analyses against simpler models suggest that CADRE's attributes yield improved generalization.
Moreover, the authors show how attention weights could be used to identify biomarker genes, partially addressing the notorious black-box property associated with many DL models.
In AGMI, attentions are used for aggregating heterogeneous feature types, including raw multiomics as well as engineered features via protein-protein interaction (PPI), gene pathways, and gene correlations with PCC (Pearson correlation coefficient), contributing to improved generalization.
Overall, papers report the predictive benefits of attentions \cite{zuo_swnet_2021, liu_graphcdr_2021, ruiwei_agmi_2021, tao_predicting_2020, jin_hidra_2021, pu_canceromicsnet_2022}, while a few papers also explore attention weights for interpretability \cite{tao_predicting_2020, jin_hidra_2021}.


Transformer is an attention-based architecture that have recently been explored in DRP models for encoding drug representations \cite{chu_graph_2022, jiang_deeptta_2022}.
In GraTransDRP \cite{chu_graph_2022}, the authors propose to extend an existing model, GraOmicDRP \cite{nguyen_graph_2021}, by modifying the drug subnetwork to include graph attention network (GAT), graph isomorphism network (GIN), and a graph transformer which learns from graph data.
In DeepTTA \cite{jiang_deeptta_2022}, a transformer module encodes drug information represented as text data (ESPF substructures \cite{huang_explainable_2019}).
Both models report significant improvement in generalization thanks to transformer modules.
Transformers have shown immense success first in language and later vision applications, and expected gain further attention from the DRP community.

\subsubsection{Graph neural network layers} \label{sec:gnn}
Designed to efficiently learn from graph data, GNNs have been extremely popular in applications where information can be represented as graphs.
With at least 14 GNN-based DRP models published since 2020 (Table S1), it is apparent that the use of GNNs with graph data becomes an emerging alternative to some of the more traditional approaches discussed above.
In DRP papers, a graph, $G = (V, E)$, is often characterized by a set of nodes $V$ (a.k.a vertices), a set of edges $E$ (a.k.a links), an adjacency matrix $W$ which contains values representing associations between the nodes, and an attribute matrix $A$ containing feature vectors representing node attributes \cite{yi_graph_2022}.
A primary challenge that researchers are facing when designing GNN-based models lies in how to transform components of the DRP problem into graphs.
Luckily, accumulated knowledge and methodologies allow viewing biological and chemical systems as networks.

Due to substantial progress in applying GNNs to drug discovery and development, \cite{gaudelet_utilizing_2021, Ramsundar-et-al-2019}, it was relatively straightforward to adopt similar techniques to DRP. Papers propose models that convert drug SMILES into graphs and utilize modern GNN layers to encode latent drug representations \cite{liu_deepcdr_2020, nguyen_graph_2021, zuo_swnet_2021, chu_graph_2022, nguyen_integrating_2022, ma_dualgcn_2022}.
The common approach is constructing a unique graph per drug with nodes and edges representing atoms and bonds, respectively, where node features are the properties of each atom. The subnetworks consists of several (usually 3 to 5) GNN layers such as GCN and GIN.

Another approach is constructing graphs using biological information of cancer samples where genes are the graph nodes and gene relationships are the edges.
There is slightly more diversity of approaches in this space as compared to drugs partially due to the use multiomic data which allow to construct graphs with heterogeneous node attributes.
The multiomics can be utilized to encode gene relationships and attributes using one or more data modalities, including correlations between genes 
\cite{ahmed_network-based_2020, ruiwei_agmi_2021, peng_predicting_2022, pu_canceromicsnet_2022}, known protein interactions (\ie, PPI) \cite{kim_graph_2021, zhu_tgsa_2021, ruiwei_agmi_2021, pu_canceromicsnet_2022} using STRING database \cite{szklarczyk_string_2021}, and relationships based on known gene pathways \cite{ruiwei_agmi_2021} using GSEA dataset \cite{subramanian_gene_2005}.
Recently, novel approaches have been explored such as heterogeneous graphs where both cell lines and drugs are encoded as graph nodes \cite{liu_graphcdr_2021, zhu_tgsa_2021, peng_predicting_2022}, and a model that utilizes diverse data types for building graphs, including differential gene expressions, disease-gene association scores and kinase inhibitor profiling \cite{pu_canceromicsnet_2022}. Despite the diversity of the different approaches, most papers report superior performance as compared baseline models.

\subsection{Learning schemes} \label{sec:dl_learning_schemes}
Various learning schemes have been proposed to improve drug response prediction in cancer models.
These techniques can be used in different configurations usually regardless of the fundamental NN building blocks that are used to construct the architecture.

\subsubsection{Autoencoders} \label{sec:ae}

When considering the large feature space of cancer and drug representations (Sections \ref{sec:tissues_rep} and \ref{sec:drugs_rep}) and the size of drug response datasets, the number of input features outnumbers the number of response samples which potentially leads to model overfitting \cite{adam_machine_2020}.
Learning predictive representations with high-dimensional data manifests a primary strength of multi-layer NNs.
A prominent example are autoencoders (AEs) which are NNs that are trained to compress and then reconstruct data in an end-to-end unsupervised learning fashion.
Given that an AE can reliably reconstruct the input data, the compressed latent representation is characterized by reduced data redundancy which improves the feature to sample ratio and can be subsequently used as inputs in downstream drug response prediction task.
The primary application of AEs in DRP models is dimensionality reduction. The compressed representation is used as input to a DRP model which is usually a NN or in some cases a classical ML model.
DeepDSC reduces gene expressions from about 20,000 genes down to 500 and concatenates the downsampled data with drug FPs as inputs to a FC-NN \cite{li_deepdsc_2021}.
VAEN exploits variational AEs (VAEs) to learn a low-dimensional representation of gene expressions which are fed into ElasticNet \cite{jia_deep_2021}.
DEERS compresses cancer features (gene expression, mutation, and tissue type) and drugs features (kinase inhibition profiles) into ten dimensions which are concatenated for a FC-NN \cite{koras_interpretable_2021}. 
Dr.VAE learns pre- and post-treatment embeddings of gene expressions and leverages those embeddings for DRP \cite{rampasek_drvae_2019}.
AutoBorutaRF combines AEs with a feature selection method followed by Boruta algorithm and RF \cite{xu_autoencoder_2019}. 
Ding et al. utilized learned representations of multiple hidden layers of the encoder as input features rather than using just the latent representation \cite{ding_precision_2018}.

AEs have also been utilized with external datasets as part of model pretraining.
DeepDR uses gene expressions and mutations from the TCGA database to train separate AEs for each omic type \cite{chiu_predicting_2019}. The pre-trained encoders are extracted from the individual AEs, concatenated, and passed as inputs to a FC-NN. The combined model was trained to predict drug response in cell lines. 
Another model uses separate AEs to encode cancer and drug features \cite{dong_variational_2021}. A GeneVAE encodes gene expressions from CCLE dataset and a Junction-Tree VAE (JT-VAE) \cite{jin_junction_2018} encodes drug molecular graphs from the ZINC database. Similar to DeepDR, the pre-trained encoders are concatenated and combined for DRP.

Dimensionality reduction and model pre-training are common applications of AEs. Yet, AEs have been utilized also in less conventional ways. 
Xia et al. proposed UnoMT, a multi-task learning (MTL) model that predicts drug response in cell lines via late-integration of multiomic features and drug representations \cite{xia_cross-study_2022}. In addition to the primary DRP task, the model predicts several auxiliary tasks where one of them is a decoder that reconstructs gene expressions.
TUGDA exploits advanced learning schemes such as domain adaption and MTL to improve generalization in datasets with limited sample size \cite{peres_da_silva_tugda_2021}. Each prediction task in the MTL corresponds to a different drug. With the goal mitigate the influence of unreliable tasks 
and reduce the risk of negative transfer, a regularization AE takes the model output and reconstructs an intermediate hidden layer. 
\subsubsection{Transfer learning} \label{sec:transfer_learning}

Transfer learning refers to learning schemes designed to improve generalization performance in a target domain $T$ by transferring knowledge acquired in a source domain $S$ \cite{pan_survey_2010}.
The most common scenario is to transfer learned representations from a source domain with large amounts of data into a target domain which suffers from insufficient data.
Transfer learning with DL have shown remarkable success in various applications \cite{zhuang_comprehensive_2021} and have recently been applied to DRP \cite{zhu_ensemble_2020, sharifi-noghabi_aitl_2020, sharifi-noghabi_drug_2021, peres_da_silva_tugda_2021, ma_few-shot_2021, prasse_pre-training_2022}.

Zhu et al. proposed an ensemble transfer learning (ETL) that extends the classical transfer learning by combining predictions from multiple models, each trained on a cell-line dataset and fine-tuned on a different cell-line dataset.
Considering three application scenarios, a NN with late-integration of gene expression and drug descriptors outperformed baselines in most cases while the LightGBM showed superior performance in certain experimental settings.
%
AITL was proposed to improve generalization across biological domains \cite{sharifi-noghabi_aitl_2020}.
The model was trained using gene expressions and drug responses from source and target domain data, where abundant cell line data serve as source and the relatively scarce data from either PDXs or patients serves as target.
A primary component for accomplishing knowledge transfer is a feature extractor subnetwork that learns shared representations using source and target features, which are passed to multi-task subnetwork for response prediction of source and target samples.
%
Ma et al. proposed TCRP, a two-phase learning framework with meta-learning as a pre-training step followed by a few-shot learning for context transfer \cite{ma_few-shot_2021}.
With the goal to obtain transferable knowledge, a meta-learning is applied iteratively by training a simple NN with different subsets of cell lines.
The same NN was trained to predict drug response in both phases but with different data.
With a few samples from the target context, the pre-trained NN was further trained for a single iteration, demonstrating performance with patient-derived cell line (PDTC) and PDXs.

ETL and AITL can be categorized as inductive transfer learning where labels from both the source and target data are used in improve generalization on target data \cite{zhuang_comprehensive_2021}.
However, abundance of per-clinical and clinical data are not labeled, and therefore, remain unused with classical transfer learning.
Velodrome is a semi-supervised model that exploits labeled cell lines and unlabeled patient data to improve out-of-distribution (OOD) generalization on datasets that remained unused during training. Generalization has been demonstrated with PDX, patient, and cell lines from non-solid tissue types.
TUGDA is an MTL and domain adaptation model which relies on cell line data and advanced learning schemes to improve predictions in data-limited cancer models \cite{peres_da_silva_tugda_2021}.
The model takes gene expressions and learns a latent shared representation that is propagated to predict drug response in a multi-task fashion with each task corresponding to a different drug. 
To enable domain adaptation from cell lines to other cancer models such as PDXs and patients, adversarial learning is used where a discriminator is branched from the shared layer to classify the type of cancer model. The discriminator is employed in supervised and unsupervised steps, with source- and target domain samples, respectively.

\subsubsection{Multi-task learning} \label{sec:mtl}

As transfer learning and multi-task learning (MTL) are related approaches \cite{pan_survey_2010, zhuang_comprehensive_2021}, several models have integrated both methods in their models \cite{sharifi-noghabi_aitl_2020, peres_da_silva_tugda_2021, xia_cross-study_2022}. However, utilizing MTL without transfer learning is also common.
%
DrugOrchestra is an original work where multiple heterogeneous data sources were curated to develop a multi-task model for simultaneously learning to predict drug response, drug target, and side effects while utilizing advanced weigh sharing approaches \cite{jiang_drugorchestra_2020}.
%
SWnet proposed a gene weight layer which scales the contributions of mutations when combined with gene expressions as a multiomic input \cite{zuo_swnet_2021}.
In its simplest form, the weight layer is trained for the entire dataset.
The authors also explored a weight matrix where each vector is determined for a single drug.
Considering $N$ drugs in the dataset, the model was trained in an MTL fashion with $N$ prediction tasks, one task for each drug, outperforming the single weight layer configuration. 
%
Zhu et al. used MTL in a pre-training step to accumulate chemical knowledge by training a GNN to predict more than a thousand biochemical properties of molecules \cite{zhu_tgsa_2021}, a strategy adopted from another work \cite{hu_strategies_2020}. The pre-trained model was integrated into the complete DRP model called TGSA.

\section{Model evaluation and comparison} \label{sec:eval_methods}
\begin{figure}[h]
    \centering
    \includegraphics[width=0.9\textwidth]{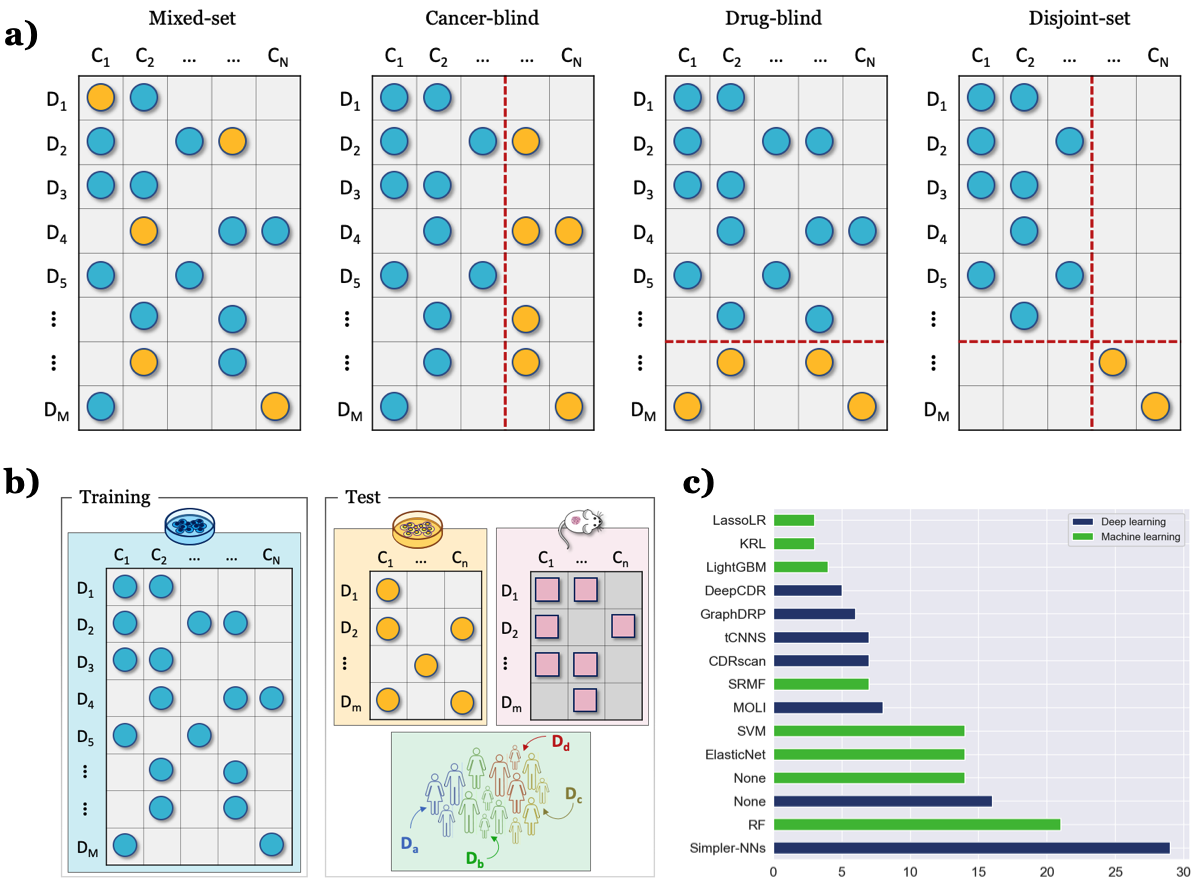}
    \caption{
    Evaluation and comparison for drug response prediction models.
    a) Data splitting strategies for evaluating performance with a single drug screening study include mixed-set, cancer-blind, drug-blind, and disjoint-set.
    b) Cross-dataset evaluation where drug response data used for training and evaluation come from different studies. 
    c) Histogram of the top fifteen most popular baseline models that were used to benchmark prediction performance of drug response prediction models. The baselines are color-coded as either deep learning (DL) or classical machine learning (ML).Simpler-NN refers to simpler versions of proposed models. The label None refers to cases where this type of baseline (\ie, ML or DL) was not used in performance analysis at all (\eg, the blue bar for None shows that 16 of the papers have not used DL-based baselines in their analysis).
    }
    \label{fig:eval_methods}
\end{figure}

Papers proposing DRP models explore advanced approaches for data representation, NN architecture design, and learning schemes. To motivate the increasing complexity of methods, it is essential to demonstrate their utility for cancer medicine and advantages against existing approaches (\ie, baseline models).
Model performance in these papers is primarily assessed via prediction generalization, \ie, prediction accuracy on a test set of unseen data samples, with several common metrics used for regression, classification, and ranking.
The choice of the test set is critical as it may illustrate the potential utility of the model for a given application in cancer medicine and significantly affect the observed performance.
To make a comparison between proposed and existing models compelling
, consistent evaluations should be utilized in terms of drug response datasets, training and test sets, evaluation metrics, and optimization efforts (\eg, HP tuning). 
This section presents common evaluation approaches used for prediction generalization in terms of drug screening datasets (single or multiple datasets), data splitting strategies (train and test), and baselines models (Fig. \ref{fig:eval_methods}).


\subsection{Single-dataset evaluation} \label{sec:single_study}
A single drug screening study is most commonly used to develop and assess performance of DRP models.
Regardless of specific cancer models the screening data space can be characterized by the involved drugs and cancers. The 2-D matrix in Fig. \ref{fig:screening_data} illustrates such a space composed of a finite number of known cancer cases and drugs, where the marked coordinates symbolize that treatment responses are available for these combinations.
Considering this 2-D space, four data splitting strategies are commonly used where each can imitate a different application scenario of drug response prediction (Fig. \ref{fig:eval_methods}a).

\subsubsection{Known cancer and drugs (mixed-set)}
To ensure a valid assessment of prediction generalization, training and test sets should contain unique cancer-drug pairs without overlap.
The overlap may occur, however, on cancer or drug (but not both) in which case a model encounters drug or cancer features during testing that also showed up during the training phase.
For example, consider two drug response samples $r_{11}=\{d_1, c_1\}$ included in the training set and $r_{12}=\{d_1, c_2\}$ in the test set.
While the pairs are unique, the drug features of $d_1$ show up during both training and testing.
The cancers or drugs that show up in test and training set are deemed as \textit{known} or \textit{seen}.
This splitting strategy is also known as mixed-set where drugs and cancers are mixed in training and test sets.
Randomly splitting cancer-drug pairs naturally results in a mixed-set analysis where drug and cancer features can show up in both training and test sets.
Due to the overlap on both feature dimensions, this analysis naturally leads to the highest performance and it is also the most common and straightforward to implement.
Because of the simplicity of this strategy almost all papers include it in their analysis.
An application of prediction models developed with the mixed-set validation is repurposing of known anticancer drugs for unexplored cancer conditions that have not been treated with these drugs.
When considering the 2-D space of known cancer cases and drugs in a given dataset, not all cancer cases and drugs have been screened against each other.
If a DRP model exhibits high generalization, it can be deployed as a virtual screening tool which helps in identifying highly sensitive cancer-drug pairs and thereby assisting in the design of experiments or treatments.

\subsubsection{Unknown cancers and known drugs (cancer-blind)} 

When a dataset is split such that an overlap occurs only on drugs but not on cancer cases, it is said that drugs are known but the cancer cases are unknown with respect to the test set.
This splitting strategy is also sometimes referred as cancer-disjoint, cancer-blind, unseen-cancer, and a special case called leave-n-cancers-out where a certain number of cancers is excluded from training set.
Splitting a drug response dataset this way requires extra work which is, therefore, less common than random splitting but still prevalent especially in the recent years.
The performance is usually worse than in the mixed-set analysis due to increased complexity in generalizing across the omics feature space of cancers.
Alternatively, this type of analysis is perhaps the most adequate in simulating the utility of DRP models for the design of personalized cancer treatment.
When considering a potential workflow of utilizing a DRP model for personalized cancer treatment, the patient tumor data is not expected to be included in the training set that was used for model development.
However, the drug treatments would be chosen from a collection of known compounds (approved or investigational) that may have been screened against other cancer cases, the drug features of which were used to train the prediction model during model development.
Models that exhibit high generalization in this scenario are considered more suitable for designing personalized cancer treatment.

\subsubsection{Known cancers and unknown drugs (drug-blind)} 

Analogously to cancer-blind analysis, the drug-blind refers to data partitioning were an overlap between the training and test sets may occur on cancers but not on drugs.
This data splitting strategy is also sometimes referred as drug-disjoint, unseen-drugs, and a special case called leave-n-drugs-out where a certain number of drugs is excluded from training set.
Interestingly, the generalization in drug-blind analysis as evaluated by common performance metrics is significantly worse than in mixed-set and cancer-blind analysis and, in some cases, models completely fail to make effective predictions. 
The worse performance can be attributed to the immense chemical space of drug compounds which imposes a challenge on predictive models of learning generalizable feature embeddings with just a few hundreds of drugs that were screened in typical drug screening studies.
In addition, it has been shown empirically that drug diversity contributes to a majority of response variation \cite{zhu_enhanced_2020, xia_cross-study_2022}, which can explain the performance drop in drug-blind analysis.
Models that excel in generalizing for unknown drugs can be useful for repurposing of non-cancer therapies to cancer indications and development of novel drugs for cancer treatment. Existing drugs approved for non-cancer diseases can be repurposed for cancer treatment, thereby decreasing expenses and speeding up the time to market.
A DRP model, exhibiting high performance in drug-blind scenario, can be utilized for in silico drug screening across cancer types and libraries of approved compounds.




\subsubsection{Unknown cancers and drugs (disjoint-set)}
By extending the drug-blind and cancer-blind analysis to both dimensions, generalization analysis can be performed where both drugs and cancers remain disjoint between training and test sets.
This analysis exhibits the worst generalization performance.
It is primarily used to assess the capacity of models to generalize in this challenging scenario where the application of such model in clinical or preclinical setting is not essentially obvious.
Models exhibiting significantly better performance as compared to baselines could be interesting cases for further exploration of the model architecture to scenarios more relevant in pre-clinical and clinical settings.



\subsection{Cross-dataset evaluation} \label{sec:multiple_studies}

Another validation scheme is to assess generalization across datasets where response data for model development and model testing are derived from different drug screening studies (Fig. \ref{fig:eval_methods}b).
The source dataset ($D_S$), used for model development, and the target dataset ($D_T$), used for evaluation, can be of the same or different cancer models.
Several selected works have been summarized in \cite{sharifi-noghabi_drug_2021}, showing that cell lines constitute the primary cancer model for $D_S$ while cell lines, PDX, and patient response data are all common options for $D_T$. 
Cross-dataset analysis is less prevalent as compared to single-dataset analysis. Primary challenges are related to data preparation and arise when standardizing metadata across datasets, including annotations of drugs, cancer samples, and omics features, as well as utilizing consistent data preprocessing steps for feature normalization (\eg, addressing batch effect) and computation of response metrics.

Inconsistencies in drug screening data across cell line studies, partially due to different experimental setups, are acknowledged reality which impedes naive integration of data from multiple sources into a single dataset \cite{haibe-kains_inconsistency_2013, gupta_normalized_2020}.
A large-scale empirical study focused on assessing generalization of DL and ML models across five cell line datasets suggests that generalization may depend on multiple attributes of $D_S$ and $D_T$ datasets, including the number of unique drugs and cell lines in $D_S$ and $D_T$, the drug and cell line overlap between $D_S$ and $D_T$, and the sensitivity assays used for measuring drug response.
Although this type of analysis requires adequate knowledge and substantial effort in data preparation and model development, demonstrating the ability to make accurate predictions across datasets or cancer models can be essential for a DRP model to become a favorable candidate for preclinical and clinical studies.

\subsection{Baselines} \label{sec:baselines}
Proposing novel DRP models could be seen unnecessary unless it leads to better performance as compared to existing models.
In predictive modeling, baseline models usually refer to existing models that serve as a point of reference for proposed models. From the collection of papers in Table S1, all models use one or more baselines of their choice as shown in Fig \ref{fig:eval_methods}c which include ML-based (green) and DL-based (blue) models.

Fig \ref{fig:eval_methods}c shows the number of times each model was used as a baseline across papers.
The most popular choice for baselines in these papers are \textit{simpler} versions of the proposed models, where simplicity is defined in the context of a given NN architecture. For example, in the case of MOLI which advocates for multi-omics late integration with triplet loss function, simpler versions of the model exploit combinations of single- instead multi-omics, early instead of late integration, and binary cross-entropy loss function instead of triplet loss function \cite{sharifi-noghabi_moli_2019}.
Using this type of baselines can be both straightforward and insightful especially with complex model architectures. On one hand, modifying existing architectures should often involve minor changes of little complexity. On the other hand, ablation analysis where NN modules are replaced with more trivial alternatives can clearly demonstrate the necessity of a proposed architecture which includes the added NN components.

The second most popular baseline model is random forest (RF) presumably due to its availability, simplicity, and predictive power. Boosting algorithms such XGBoost and LightGBM are less prevalent although poses similar characteristics as RF yet faster and often more predictive.
The label \textit{None} refers to cases where this type of baseline (\ie, ML or DL) was not used in performance analysis at all (\eg, the green bar for \textit{None} shows that 16 of the papers have not used DL-based baselines in their analysis).

The details of how models are compared are a critical aspect. 
We observe that a common approach for comparing models with baselines is to use performance scores from the original publications without actually implementing the baselines and benchmarking performance on the same dataset.
Model performance is likely to depend on feature types and specific samples allocated for training and testing. Therefore, using reported scores makes sense in cases where benchmark datasets have been established with clearly designated training and test samples. However, this is not the case in drug response prediction tasks.
Due to lack of consensus, reproducing results from DRP papers is challenging \cite{chen_how_2021}.
Alternatively, there are papers that diligently reproduce models from other papers and use these as baselines to benchmark the performance of models.
Despite the extra amount of work, such analysis have higher chances to establish fidelity and attract the community.

\section{Discussion} \label{sec:discussion}
In recent years, AI approaches have taken a notable place in health care \cite{bohr_chapter_2020}.
Their promise of utilizing large data arrays 
to improve treatment protocols, aid drug development efforts, and increase diagnostic precision is attractive to the clinical research community.
However, navigating a complex landscape of factors affecting clinical outcomes is a challenging task.
We can see it in the number of emerging papers on the DRP problem \cite{clyde2022ai}. The research community put in monumental efforts to develop methods that leverage multiomics and drug data, explore the applicability of the existing methods, and bridge the gap between results on ubiquitously used cell lines and patient-derived models.
However, this area is facing multiple challenges related to the lack of unified evaluation framework, generalizability, interpretability as well as challenges relating to computational representations of omic data, biological response, and drug representations.

\textbf{Unified framework for model evaluation and comparison}.
Currently, there is no standard or accepted framework for evaluation or comparison of cancer drug response models.
Model development and performance comparison against baselines is frequently accomplished with different compositions of datasets, inconsistent training and testing splits, and diverse scoring metrics, using varying protocols for hyperparameter optimization or none at all.
Yet, the majority of papers report to outperform current state-of-the-art (SOTA) in the task of drug response prediction.
In ML, SOTA refers to the best model for a specific task as evaluated by concrete performance criteria on a benchmark dataset of predetermined test set samples. 
Thousands of benchmark datasets and prediction tasks are publicly available for different applications ranging from vision and language to drug discovery and tumor segmentation.
Whereas most papers follow the same general model development workflow (Fig. \ref{fig:drp_schematic}), benchmark datasets and agreed-upon evaluation criteria have yet been established and adopted in the community.
It makes identifying the most promising research areas challenging and impedes the research community from making directed efforts to breach them.
Creating an ImageNet moment for DRP via established benchmark datasets, consistent test sets, robust evaluation criteria, and a platform for publishing and monitoring SOTA models should be on the critical path for our community.

\textbf{Generalizability}.
Despite the breadth of methods, the majority of papers have focused on improving predictions in cell lines, often demonstrating only a marginal improvement in prediction generalization.
While cell lines remain a primary biological media to study cancer and conduct drug screenings, the potential utility of prediction models for improving patient care is not immediately evident, and several questions naturally arise. How well models trained with cell lines would generalize to xenografts or patients? How much one could rely on any given model prediction in a decision making process? What are the potential clinical application for these models?
Addressing questions important from the user perspective and suggesting use case scenarios have not been the driving mechanism behind the majority of published models.
Yet, certain trends aiming to respond to these challenges emerge: transfer learning that utilize abundant cell line datasets to improve predictions in PDX and patients, uncertainty quantification allowing to estimate the confidence for each model prediction \cite{peres_da_silva_tugda_2021}, and ranking learning models principally suitable for personalized treatment recommendation \cite{prasse_matching_2022, su_srdfm_2022, liu_deep_2022}.
Demonstrating the utility of DRP models, integrated in a larger patient care workflow, could provide the needed user-centric view and navigate the community towards developing application-aware models.

\textbf{Interpretability}.
Despite the demonstrated success of the DL on expansive datasets, delegating the decision-making process on patient well-being to the black box is widely contested \cite{naik2022legal}.
Thus, providing not only precise but also salient results is a task of paramount importance for the clinical community.
To address this challenge, model-agnostic methods such as Integrated Gradients \cite{sundararajan_axiomatic_2017} and SHAP \cite{lundberg_unified_2017}, designed to explain model predictions, have been applied post-training to DRP models \cite{snow_interpretable_2021, tang_explainable_2021, prasse_pre-training_2022}, where the explanation is provided in the form of most important features attributing to model predictions.
However, it has been shown that these methods can lead to highly misleading information in applications more comprehensible to humans such as image classification, and therefore, attempting to explain black boxes with post hoc methods is perceived as dangerous in high-stakes decision-making domains \cite{rudin_stop_2019}.
Instead of explaining predictions, the alternative is to design interpretable models that are understandable by domain experts or provide insights into the decision-making process \cite{murdoch_definitions_2019, barredo_arrieta_explainable_2020}.
Unfortunately, no single definition exists, and identifying a model as interpretable is considered domain-specific \cite{rudin_stop_2019, murdoch_definitions_2019}.
Several papers referring to their DRP models as interpretable integrate domain structural knowledge into the model form \cite{deng_pathway-guided_2020, kuenzi_predicting_2020, zhang_predicting_2021, snow_interpretable_2021}.
Currently, no clear definition of an interpretable DRP model yet exists nor the extent to which it might improve cancer treatment is known.
Therefore, it is challenging to evaluate whether existing efforts in this direction are in line with the views of the clinical community at large on this important matter.
Addressing the interpretability issue will perhaps require better framing of what interpretable DRP models are and how much performance drop could be tolerated, if any, for improved interpretability. 
Collaboration with cancer biologists and clinical oncologists would be essential to advance this direction.

\textbf{Variability and high dimensionality associated with omics data}.
Omics values differ substantially depending on the underlying biological model (\eg, cell line, organoid, xenograft), experimental protocols and selected platforms (\eg, Illumina, Nanopore), technical variations (batch effect), and computational pipeline for processing raw data.
Identifying the potential sources of variation and taking directed measures to mitigate the undesired differences is critical, especially in cross-domain generalization scenarios where the similarity between the source and target domains is a fundamental assumption \cite{sakellaropoulos_deep_2019, sharifi-noghabi_aitl_2020, sharifi-noghabi_out--distribution_2021, peres_da_silva_tugda_2021}.
A major conclusion of the NCI-DREAM challenge was that gene expression modality provides most of the predictive power for cell line drug sensitivity prediction with additional improvement when combined with other omics types \cite{costello_community_2014}.
This collaborative effort contributed to the adoption of gene expression (GE; \eg, microarrays or RNA-Seq) or combining them with other omics in DRP models.
Integrating multiomics into the learning process further exacerbates the already problematic feature size of single omics data. 
A few papers indeed demonstrate significant performance boost with multiomics \cite{liu_graphcdr_2021, tang_explainable_2021, ma_dualgcn_2022} but the majority report only marginal improvement \cite{chiu_predicting_2019, zhu_tgsa_2021, park_superfelt_2021, liu_graphcdr_2021, chu_graph_2022, liu_deep_2022}.
As opposed to DREAM participators which utilized agreed-upon datasets and scoring metrics, the DRP models in Table S1 
substantially differ among them as discussed earlier, largely contributing to discrepancies and mixed conclusions.
Analysis across multiple models and omics types is required to evaluate the predictive capabilities of individual data types and their subsets and make unbiased and coherent conclusions.
Such analysis should incorporate recent trends which encode biological information such as protein-protein interactions (PPI), gene correlations, and pathway information \cite{tang_explainable_2021, ruiwei_agmi_2021, pu_canceromicsnet_2022, ma_dualgcn_2022}.

\textbf{Representation of biological treatment response}.
In vitro response data (\eg, cell line and organoid) is usually derived from a series of inherently noisy cell viability experiments and subsequent curve fitting.
However, these curves do not always have a good fit, and usage of a single point on the fitted curve may result in a substantial information loss.
While the use of IC50 as a prediction variable has been the most prevalent in regression models (Fig. \ref{fig:rsp_cell}), recent assessment studies could shift this trend towards the use of global and generally more robust measures such as AUC and AAC \cite{sharifi-noghabi_drug_2021, xia_cross-study_2022}.
With the objective of utilizing prediction models in decision-making situations, another option is converting continuous into discrete values.
This is particularly common in cross-domain generalization from in vitro to in vivo where in vivo response is generally encoded as discrete values.
The third option is generating a rank list of items. This has been applied to a personalized treatment recommendation given cancer information. Surprisingly, despite the success of recommendation algorithms and the direct relevance to precision oncology, only few methods have been explored \cite{su_srdfm_2022, prasse_matching_2022, liu_deep_2022}.

\textbf{Vastness of chemical space and its representations}.
Reviewed papers consistently report the response prediction accuracy drop for compounds, previously unseen by the model.
Inability to reliably overcome this limitation severely limits the practical value of DRP models to virtual drug screening or drug design applications.
Representing drug molecules with graphs and using GNN for learning response prediction has recently inspired many DRP architectures.
The use of graph structures is motivated by the proposition that molecular graphs better capture intrinsic chemical properties of molecules \cite{liu_graphcdr_2021, ma_dualgcn_2022, deep_surrogate_docking}.
Whereas this is presumably expected to produce a better prediction of drug response, we have not found a comprehensive study that could assert this hypothesis.
In fact, in a related field of molecular property prediction, a comparison with multiple datasets and prediction tasks suggests that on average models that use FPs or descriptors outperform graph-based models \cite{jiang_could_2021}.
We have found only one study comparing side-by-side the added value of molecular graphs against SMILES, reporting less than 0.5\% improvement as evaluated by Pearson correlation coefficient \cite{yan_deep_2022}. 
As in the case of cancer representation, further studies are required across various models and datasets to assess the predictive capabilities of molecular graphs and other drug representations to DRP.
Several underexplored but interesting directions include SMILES combined with transformer-based models, kinase inhibition profiles representing kinase inhibitor therapies \cite{koras_interpretable_2021}, and 2D-CNNs that learn from FPs or descriptors transformed into images \cite{zhu_converting_2021}.

This list of challenges is in no way conclusive, as we refrained from studying even more complicated scenarios emerging in polypharmacy, even though the treatment of patients in the clinic often involves combination therapy. 
Meanwhile, most of the prediction models to date focus on single-drug treatments.
This highlights a significant disconnect between the modeling community and the current patient standard of care.
However, even this simplified problem raises many obstacles, and making progress in overcoming them would significantly benefit the scientific community because they are relevant to a much wider domain of AI-driven biological research.
Each of the highlighted problems has multiple potential solutions, and we hope to see subsequent progress in the near future, allowing AI to deliver on its grand promise.

\section*{Funding}
The research was part of the IMPROVE project under the NCI-DOE Collaboration Program that has been funded in whole or in part with Federal funds from the National Cancer Institute, National Institutes of Health, Task Order No. 75N91019F00134 and from the Department of Energy under Award Number ACO22002-001-00000.

\bibliographystyle{ieeetr}

\bibliography{main}  

\end{document}





\title{Supplementary material}
\maketitle

The supplementary material contains: 1) abbreviations aggregated by topics, and 2) Table S1 that lists the curated drug response prediction models.

\section*{Abbreviations}
\subsection*{NN modules}
NN: neural network.
1D-CNN and 2D-CNN: one- and two-dimensional convolutional NN.
Att: attention modules.
GNN: graph NN.
Inter: interpretable NN.
RNN: recurrent NN.
RC: residual connections.

\subsection*{Learning schemes}
AE: autoencoder.
AL: adversarial learning.
BNN: Bayesian NN.
CF: collaborative filtering.
CL: contrastive learning.
DeepFM: deep factorization machine.
EL: ensemble learning.
ML: meta learning.
MVL: multi-view learning.
MTL: multi-task learning.
MTSPT: multi-task supervised pre-training.
RL: reinforcement learning.
SSPT: self-supervised pre-training.
TL: transfer learning.
Tran: transformer.

\subsection*{Cancer features}
CNV: copy number variation.
GE: gene expression.
Methyl: methylation.
Mu: mutation.
RPPA: reverse phase protein arrays.
TT: tissue type.

\subsection*{Drug features}
SMILES: simplified molecular-input line-entry system.
DD: drug descriptors.
FP: drug fingerprints.
MG: molecular graphs.
KIP: drug kinase inhibition profiles.

\subsection*{Response}
AAC: area above the dose response curve.
AUC: area under the dose response curve.
Rank: ranking.
IC50-to-Bin: IC50 converted to binary.
AUC-to-MC: AUC converted to multi-class.
GI50-to-bin: GI50 converted to binary.
Bin-enc-IC50: binary encoded IC50.
Bin-enc-AUC: binary encoded AUC.
Bin-enc-AAC: binary encoded AAC.
Bin-enc-GI50: binary encoded GI50.
MC-enc-IC50: multi-class encoded IC50.
CTR: continuous tumor response.

\newgeometry{top=20mm, left=10mm}
\begin{table}[!t]\centering
\footnotesize
\renewcommand{\arraystretch}{2}
\begin{adjustbox}{angle=90}
\begin{tabular}{ 
    >{\raggedright\arraybackslash}p{0.07\textwidth} > {\raggedright\arraybackslash}p{0.07\textwidth} >{\raggedright\arraybackslash}p{0.07\textwidth} >{\raggedright\arraybackslash}p{0.07\textwidth} >{\raggedright\arraybackslash}p{0.07\textwidth} >{\raggedright\arraybackslash}p{0.07\textwidth} >{\raggedright\arraybackslash}p{0.07\textwidth} >{\raggedright\arraybackslash}p{0.07\textwidth} >{\raggedright\arraybackslash}p{0.07\textwidth} >{\raggedright\arraybackslash}p{0.37\textwidth}
}
\multicolumn{10}{c}{\normalsize Table S1: List of monotherapy drug response prediction models. \vspace{10pt}}\\\toprule
Paper &Model &Framework &Methods &Cancer features &Drug features &Response cell line &Response PDX &Response patient &Code URL \\\midrule
Yan et al. 2022 &DGSDRP &PyTorch &1D-CNN, GNN, bRNN &Mu &MG &IC50 &NA &NA &https://github.com/YanXf7/DGSDRP \\
Xia et al. 2022 &UnoMT &TF1 &RC, AE, MTL &GE &DD, FP &AUC &NA &NA &NA \\
Wang et al. 2022 &MvMo &PyTorch &MVL &GE, Methyl, Mu &MG &IC50 &NA &NA &NA \\
Su et al. 2022 &SRDFM &TF1 (w/o Keras) &DeepFM &GE &FP &Rank &NA &NA &https://github.com/RanSuLab/SRDFM \\
Pu et al. 2022 &Cancer Omics Net &PyTorch &Att, GNN &GE &KIP &GR &NA &NA &https://github.com/pulimeng/CancerOmicsNet \\
Prasse et al. 2022 &Conv NN &TF2 &1D-CNN, TL &GE &SMILES &IC50 &CTR &NA &https://github.com/prassepaul/mlmed\_transfer\_learning \\
Prasse et al. 2022 &None &TF2 (w/o Keras) &1D-CNN, Att, RC &GE &SMILES &Rank &NA &NA &https://github.com/prassepaul/mlmed\_ranking \\
Peng et al. 2022 &MOFGCN &PyTorch &GNN &CNV, GE, Mu &FP &Bin-enc-IC50 &NA &NA &https://github.com/weiba/MOFGCN \\
Nguyen et al. 2022 &Gra Omic DRP &PyTorch &1D-CNN, GNN &CNV, GE, Methyl, Mu &MG &IC50, Bin-enc-IC50 &NA &NA &NA \\
Ma et al. 2022 &DualGCN &TF1 &GNN &CNV, GE &MG &IC50 &NA &IC50-to-Bin &https://github.com/horsedayday/DualGCN \\
Liu et al. 2022 &PPORank &NA &RNN, RL &GE, Methyl &NA &Rank &NA &Rank &NA \\
Jiang et al. 2022 &DeepTTA &PyTorch &Tran &GE &FP &IC50-to-Bin &NA &NA &https://github.com/jianglikun/DeepTTC \\
Chu et al. 2022 &Gra Trans DRP &PyTorch &1D-CNN, GNN, Tran &CNV, GE, Mu &MG &IC50 &NA &NA &https://github.com/chuducthang77/GraTransDRP \\
\end{tabular}
\end{adjustbox}
\end{table}

\newgeometry{top=20mm, left=20mm}
\begin{table}[!t]\centering
\footnotesize
\renewcommand{\arraystretch}{2}
\begin{adjustbox}{angle=90}
\begin{tabular}{ 
    >{\raggedright\arraybackslash}p{0.07\textwidth} > {\raggedright\arraybackslash}p{0.07\textwidth} >{\raggedright\arraybackslash}p{0.07\textwidth} >{\raggedright\arraybackslash}p{0.07\textwidth} >{\raggedright\arraybackslash}p{0.07\textwidth} >{\raggedright\arraybackslash}p{0.07\textwidth} >{\raggedright\arraybackslash}p{0.07\textwidth} >{\raggedright\arraybackslash}p{0.07\textwidth} >{\raggedright\arraybackslash}p{0.07\textwidth} >{\raggedright\arraybackslash}p{0.37\textwidth}
}
\toprule
Paper &Model &Framework &Methods &Cancer features &Drug features &Response cell line &Response PDX &Response patient &Code URL \\\midrule
Zuo et al. 2021 &SWnet &PyTorch &1D-CNN, GNN, Att, MTL &GE, MU &FP, MG &IC50 &NA &NA &https://github.com/zuozhaorui/SWnet \\
Zhu et al. 2021 &TGSA &PyTorch &GNN, MTSPT, SSPT &CNV, GE, Mu &FP, MG &IC50 &NA &NA &https://github.com/violet-sto/TGSA \\
Zhu et al. 2021 &IGTD &TF1 &2D-CNN &GE &DD &AUC &NA &NA &https://github.com/zhuyitan/IGTD \\
Zhang et al. 2021 &Cons Deep Signaling &TF1 &Int. NN &CNV, GE &NA &AUC &NA &NA & \\
Tang et al. 2021 &PathDSP &PyTorch &FC-NN (EI) &CNV, GE, Mu &FP &IC50 &NA &NA &https://github.com/TangYiChing/PathDSP \\
Snow et al. 2021 &BDKANN+ &TF2 &Int. NN &GE &NA &AAC &NA &NA &https://github.com/osnow/BDKANN \\
Sharifi-Noghabi et al. 2021 &Velodrome &PyTorch &TL, AL &GE &NA &AAC &Bin &Bin &https://github.com/hosseinshn/Velodrome \\
Peres da Silva et al. 2021 &TUGDA &PyTorch &AE, AL, BNN, TL, MTL &GE &NA &IC50 &CTR &Bin &https://github.com/CSB5/TUGDA \\
Partin et al. 2021 &None &TF2 &FC-NN (LI) &GE &DD &AUC &NA &NA &https://github.com/adpartin/dr-learning-curves \\
Park et al. 2021 &Super. FELT &PyTorch &FC-NN (LI) &CNV, GE, Mu &NA &Bin-enc-IC50 &Bin &Bin &https://github.com/DMCB-GIST/Super.FELT \\
Nguyen et al. 2021 &Graph DRP &PyTorch &1D-CNN, GNN &CNV, Mu &MG &IC50 &NA &NA &https://github.com/hauldhut/GraphDRP \\
Malik et al. 2021 &None &Matlab &FC-NN (EI) &CNV, GE, Methyl, Mu &NA &IC50-to-Bin &NA &NA &https://github.com/TeamSundar/BRCA\_multiomics \\
\end{tabular}
\end{adjustbox}
\end{table}

\begin{table}[!t]\centering
\footnotesize
\renewcommand{\arraystretch}{2}
\begin{adjustbox}{angle=90}
\begin{tabular}{ 
    >{\raggedright\arraybackslash}p{0.07\textwidth} > {\raggedright\arraybackslash}p{0.07\textwidth} >{\raggedright\arraybackslash}p{0.07\textwidth} >{\raggedright\arraybackslash}p{0.07\textwidth} >{\raggedright\arraybackslash}p{0.07\textwidth} >{\raggedright\arraybackslash}p{0.07\textwidth} >{\raggedright\arraybackslash}p{0.07\textwidth} >{\raggedright\arraybackslash}p{0.07\textwidth} >{\raggedright\arraybackslash}p{0.07\textwidth} >{\raggedright\arraybackslash}p{0.37\textwidth}
}
\toprule
Paper &Model &Framework &Methods &Cancer features &Drug features &Response cell line &Response PDX &Response patient &Code URL \\\midrule
Ma et al. 2021 &TCRP &PyTorch &TL, ML &GE, Mu &NA &AUC &CTR &NA &https://github.com/idekerlab/TCRP \\
Liu et al. 2021 &Graph CDR &PyTorch &1D-CNN, Att, GNN, CL &GE, Methyl, Mu &MG &Bin-enc-AAC, Bin-enc-IC50 &NA &NA &https://github.com/BioMedicalBigDataMiningLab/ GraphCDR \\
Koras et al. 2021 &DEERS &PyTorch &AE &GE, Mu &KIP &AUC, IC50 &NA &NA &https://github.com/kkoras/rec-system-for-drug-response \\
Kim et al. 2021 &DrugGCN &TF1 (w/o Keras) &GNN &GE &NA &AUC, IC50 &NA &NA &https://github.com/BML-cbnu/DrugGCN \\
Jin et al. 2021 &HiDRA &TF1 &Att &GE & &IC50 &NA &NA &NA \\
Jiang et al. 2021 &Drug Orchestra &PyTorch &MTL &GE & &IC50 &CTR &NA &https://github.com/jiangdada1221/DrugOrchestra \\
Jia et al. 2021 &VAEN &TF1 &VAE &GE &NA &AAC, IC50 &NA &AA-to-Bin &https://github.com/bsml320/VAEN/ \\
Feng et al. 2021 &AGMI &PyTorch &Att, GNN &CNV, GE, Mu &MG &IC50 &NA &NA &NA \\
Emdadi et al. 2021 &Auto-HMM-LMF &NumPy &AE &CNV, GE, Mu, TT &FP &Bin-enc-IC50 &NA &NA &https://github.com/emdadi/Auto-HMM-LMF \\
Dong et al. 2021 &None &PyTorch &JT-VAE, VAE &GE &MG &IC50 &NA &NA &https://github.com/JIAQING-XIE/Machine-Learning-in-Genomes/tree/main \\
Zhu et al. 2020 &tDNN &TF1 &TL, EL &GE &DD &AUC &NA &NA &NA \\
Tao et al. 2020 &CADRE &PyTorch &Att, CF &GE &NA &Bin-enc-AAC &NA &NA &https://github.com/yifengtao/CADRE \\
Sharifi-Noghabi et al. 2020 &AITL &PyTorch &TL, MTL &GE &NA &IC50 &Bin &Bin &https://github.com/hosseinshn/AITL \\
\end{tabular}
\end{adjustbox}
\end{table}

\begin{table}[!t]
\footnotesize
\renewcommand{\arraystretch}{2}
\begin{adjustbox}{angle=90}
\begin{tabular}{ 
    >{\raggedright\arraybackslash}p{0.07\textwidth} > {\raggedright\arraybackslash}p{0.07\textwidth} >{\raggedright\arraybackslash}p{0.07\textwidth} >{\raggedright\arraybackslash}p{0.07\textwidth} >{\raggedright\arraybackslash}p{0.07\textwidth} >{\raggedright\arraybackslash}p{0.07\textwidth} >{\raggedright\arraybackslash}p{0.07\textwidth} >{\raggedright\arraybackslash}p{0.07\textwidth} >{\raggedright\arraybackslash}p{0.07\textwidth} >{\raggedright\arraybackslash}p{0.37\textwidth}
}
\toprule
Paper &Model &Framework &Methods &Cancer features &Drug features &Response cell line &Response PDX &Response patient &Code URL \\\midrule
Liu et al. 2020 &DeepCDR &TF1 &1D-CNN, GNN &GE, Methyl, Mu &MG &IC50 &NA &Bin &https://github.com/kimmo1019/DeepCDR \\
Li et al. 2020 &MFNN &TF1 &1D-CNN, RNN &GE, DD &DD &IC50 &NA &NA &NA \\
Kuenzi et al. 2020 &DrugCell &PyTorch &Int. NN &Mu, FP &FP &AUC &AUC-to-Bin &AUC-to-Bin &https://github.com/idekerlab/DrugCell \\
Deng et al. 2020 &pathDNN &PyTorch &Int. NN &GE &NA &AAC &NA &NA &http://pathdnn.denglab.org \\
Daoud et al. 2020 &Q-Rank &R &RL &GE, Methyl, Mu, RPPA &DD &GI50-to-Bin &NA &NA &https://github.com/salma2018/Q-Rank \\
Choi et al. 2020 &RefDNN &TF1 (w/o Keras) &FC-NN (EI) &GE &FP &Bin-enc-IC50 &NA &NA &https://github.com/mathcom/RefDNN \\
Bazgir et al. 2020 &REFINED &TF2 &2D-CNN &GE &DD &IC50, GI50, Bin-enc-GI50 &NA &NA &https://github.com/omidbazgirTTU/REFINED \\
Ahmed et al. 2020 &None &PyTorch &GNN &GE &NA &AUC &NA &NA &https://github.com/compbiolabucf/drug-sensitivity-prediction \\
Zhao et al. 2019 &None &TF1 &FC-NN (EI) &GE &NA &AAC-to-mc &NA &NA &NA \\
Xu et al. 2019 &Auto BorutaRF &H2O.ai &AE &CNV, GE, Mu &NA &Bin-enc-IC50 &NA &NA &https://github.com/bioinformatics-xu/AutoBorutaRF \\
Sharifi-Noghabi et al. 2019 &MOLI &PyTorch &FC-NN (LI) &CNV, GE, Mu &NA &Bin-enc-IC50 &Bin &Bin &https://github.com/hosseinshn/MOLI \\
Sakellaropoulos et al. 2019 &--- &H2O.ai &FC-NN (EI) &GE &NA &IC50-to-Bin &NA &Bin &https://github.com/TeoSakel/deep-drug-response \\
\end{tabular}
\end{adjustbox}
\end{table}

\begin{table}[!t]
\footnotesize
\renewcommand{\arraystretch}{2}
\begin{adjustbox}{angle=90}
\begin{tabular}{ 
    >{\raggedright\arraybackslash}p{0.07\textwidth} > {\raggedright\arraybackslash}p{0.07\textwidth} >{\raggedright\arraybackslash}p{0.07\textwidth} >{\raggedright\arraybackslash}p{0.07\textwidth} >{\raggedright\arraybackslash}p{0.07\textwidth} >{\raggedright\arraybackslash}p{0.07\textwidth} >{\raggedright\arraybackslash}p{0.07\textwidth} >{\raggedright\arraybackslash}p{0.07\textwidth} >{\raggedright\arraybackslash}p{0.07\textwidth} >{\raggedright\arraybackslash}p{0.37\textwidth}
}
\toprule
Paper &Model &Framework &Methods &Cancer features &Drug features &Response cell line &Response PDX &Response patient &Code URL \\\midrule
Rampášek et al. 2019 &Dr.VAE &PyTorch &VAE &GE &NA &Bin-enc-AAC &NA &NA &https://github.com/rampasek/DrVAE \\
Manica et al. 2019 &MCA &TF1 (w/o Keras) &1D-CNN, Att, bRNN, RC &GE &SMILES &IC50 &NA &NA &https://github.com/drugilsberg/paccmann \\
Liu et al. 2019 &tCNNS &TF1 (w/o Keras) &1D-CNN &CNV, Mu &SMILES &IC50 &NA &NA &https://github.com/Lowpassfilter/tCNNS-Project \\
Li et al. 2019 &DeepDSC &TF1 &AE &GE &FP &IC50 &NA &NA &NA \\
Joo et al. 2019 &DeepIC50 &TF1 &1D-CNN &Mu &DD, FP &mc-enc-IC50 &NA &mc &https://github.com/labnams/DeepIC50 \\
Chiu et al. 2019 &DeepDR &TF1 &AE &GE, Mu &NA &IC50 &NA &NA &NA \\
Oskooei et al. 2018 &PaccMann &TF1 (w/o Keras) &1D-CNN, Att, bRNN &GE &SMILES &IC50 &NA &NA &https://github.com/drugilsberg/paccmann \\
Ding et al. 2018 &--- &Matlab &AE &CNV, GE, Mu &NA &Bin-enc-AAC &NA &NA &NA \\
Chang et al. 2018 &CDRscan &TF1 &1D-CNN, EL &Mu &FP &IC50 &NA &NA &NA \\
Menden et al. 2013 &--- &Java &FC-NN (EI) &CNV, Mu &DD &IC50 &NA &NA &NA \\
\end{tabular}
\end{adjustbox}
\end{table}